\documentclass{article}
\usepackage{natbib}
\usepackage{amsfonts}
\usepackage{dsfont}  

\usepackage{natbib}
\usepackage{graphicx}
\usepackage{multirow}
\usepackage[centertags]{amsmath}
\usepackage{multicol}        
\usepackage{url}
\usepackage{subcaption}
\usepackage{caption}
\usepackage{float}
\floatstyle{plaintop}
\restylefloat{table}
\usepackage{booktabs}
\newtheorem{definition}{Definition} 

\title{One-class classification with application to forensic analysis}
\author{Laura Anderlucci, Francesca Fortunato, Angela Montanari \\ Department of Statistical Sciences, University of Bologna, Italy.}

\begin{document}

\maketitle


\begin{abstract} 
The analysis of broken glass is forensically important to reconstruct the events of a criminal act. In particular, the comparison between the glass fragments found on a suspect (recovered cases) and those collected on the crime scene (control cases) may help the police to correctly identify the offender(s).
The forensic issue can be framed as a one-class classification problem. One-class classification is a recently emerging and special classification task, where only one class is fully known (the so-called \emph{target} class), while information on the others is completely missing. We propose to consider classic Gini’s \emph{transvariation probability} as a measure of typicality, i.e. a measure of resemblance between an observation and a set of well-known objects (the control cases). The aim of the proposed \emph{Transvariation-based One-Class Classifier} (TOCC) is to identify the best boundary around the target class, that is, to recognise as many target objects as possible while rejecting all those deviating from this class.

\end{abstract}



\section{Introduction}

Burglaries and crime offences are frequently characterized by the breakage or the damage of some glass.
Windows smashed vigorously to force the entry and get access to private places, lamps and bottles used to hit someone or something, glass furnitures and headlamps hurt by accident, car glasses fractured by fired bullets or collisions are just a few examples of how it may happen.
As a consequence of these acts, fragments of glass scatter randomly all over the crime scene and on the offenders. In so doing, such fragments become unavoidable trace evidences and, thus, they can help the police to know more about how the crime was committed.

Usually, glass chunks arising from a breakage have a linear dimension smaller than 0.5mm; for this reason, the comparison between different fragments is often made on the basis of some analytical results: the Glass Refractive Index ($RI$), measured by instrumental methods such as m-XRF, LA-ICP-MS, SEM-EDX, and the chemical composition ($Na$, $Mg$, $Al$, $Si$, $K$, $Ca$, $Ba$, $Fe$), measured by a scanning electron microscope.

The traditional purpose of glass analysis for forensics is to evaluate whether fragments found on the suspect (\emph{recovered} cases) can be considered from the same source as those from the location at which the offence took place (\emph{control} cases)\citep{evett1987rule}.

In the forensic science literature, this issue has been already addressed within a hypothesis testing framework by using a likelihood ratio (LR) test \citep[see][]{aitken2007two}:
\begin{equation}
\label{eq:LR}
LR = \frac{f(RI, Na', Mg', Al', Si', K', Ca', Ba', Fe'|H_0)}{f(RI, Na', Mg', Al', Si', K', Ca', Ba', Fe'|H_1)}.
\end{equation}

This requires the estimation of a full model $f(\cdot|\cdot)$ for the two competing hypotheses: $H_0$, the prosecution/null hypothesis that both \emph{recovered} and \emph{control} glasses come from the same source, and $H_1$, the defence/alternative proposition that they have different origin.
In equation \ref{eq:LR} each $\cdot'$ refers to the ratio of the elemental concentration to the oxigen, $O$, one.

The problem of assessing whether the evidence is compatible with the control samples can also be framed as a \emph{one-class classification} task.
In fact, one-class classification methods aim to decide whether an object whose origin is completely unknown belongs to a particular class (the so-called ``target'' class, which, according to the terminology used before, includes the control cases only).
As no information is available on the non-target objects, one-class classification is a difficult classification problem because it has to build a precise descriptive instead of discriminant model of the target class with enough generalisation ability \citep{liu2016modular}.

In \cite{tax2001one} a detailed description of the methods for one-class classification tasks are discussed and presented. Several algorithms and methodologies have been proposed in the statistics literature so far. Major approaches can be casted into three groups: \emph{density methods}, \emph{boundary methods} and \emph{reconstruction methods}.

Procedures in the first set estimate the probability density function of the target class $\chi$, $f(x)$, with $x \in \chi$, and set a threshold, $t$, on the resulting densities; in this way a target and an outlier region can be obtained.  The density can be estimated via the most common density estimators: Parzen density estimators \citep{bishop1994novelty,tarassenko1995novelty}, Gaussian models \citep{parra1996statistical}, mixtures of Gaussians \citep{mclachlan2000finite,fraley2002model}, Kernel Density Estimation (KDE) and histograms \citep[see][for an exhaustive description]{scott2015multivariate}, $K$-nearest-neighbors (Knn) estimation \citep{ripley2007pattern}, just to name a few.
These techniques usually work very well, especially when the sample size is sufficiently large and the model assumed to describe the target distribution is appropriate. However, their actual implementation could be limited as the choice of the best model is not trivial and it requires a large number of training objects to overcome the curse of dimensionality.

Boundary methods aim to define the best boundary around the target data, avoiding a demanding estimation of the complete density. Here, the classification issue is performed by evaluating the distance of a given object from the target class and, then, by comparing it with a threshold $t$; the latter is directly derived on the distance measures and adjusted to ensure a predefined sensitivity, $s$, i.e. the proportion of target observations that are correctly identified. Boundary algorithms heavily rely on the distances between observations and, thus, they are very sensitive to the scaling of the features. In this case, although the required sample size is smaller than for density methods, the crucial task lies on the definition of appropriate distance measures. The $K$-centers algorithm \citep{ypma1998support}, the $\nu$ Support Vector Classification ($\nu$-SVC) of \cite{scholkopf2000support} and the Support Vector Data Description (SVDD) of \cite{tax2004support} represent a few examples of such class of methods. In addition to these, procedures based on the concept of data depth can be added to the set \citep[see, among others,][]{dang2010nonparametric,chen2009outlier,ruts1996computing}. In fact, statistical depth functions can be exploited to measure the ``extremeness'' or ``outlyingness'' of a data point with respect to a given data set as they provide center-outward ordering of multi-dimensional data. In one-class classification issues all the observations that significantly deviate from the data cloud are indeed expected to be more likely characterized by small depth values than large ones. Boundaty algorithms are completely data-driven and avoid strong distributional assumption; in addition, for a low dimensional input space, they provide intuitive visualization of the data set by finding peeling and depth contours (e.g. bagplot, convex hull, \dots).

Reconstruction methods are based on some assumptions about the data generating process or about the data clustering characteristics and then, describe the objects by using their {\em reconstruction error}, $\varepsilon_{reconstr}$, that is the difference between the fitted and the observed values. Since the underlying model or structure is supposed to well represent the target class, $\varepsilon_{reconstr}$ can be considered as measure of distance from $x$ to this set. Methods in this class have not been primarily derived for one-class classification purposes, but rather to simply model the data; points that do not belong to the target class are expected to be represented worse than true target objects and, therefore, their reconstruction error is supposed to be high. Among the most common reconstruction algorithms, we can find $K$-means \citep{lloyd1982least}, the Learning Vector Quantization (LVQ) by \cite{carpenter1991art}, the Self-Organizing Maps (SOM) by \cite{kohonen1998self}, Principal Component Analysis (PCA) and mixture of PCAs \citep{tipping1999mixtures} and the autoencoders by \cite{japkowicz1995novelty}.

Recent approaches include deep learning methods, such as deep neural networks, to extract common factors of variations from the data \citep{ruff2018deep} and deep support vector machines \citep{erfani2016high}.

In this paper a novel one-class classification algorithm based on Gini's transvariation probability as a measure of resemblance is introduced; the proposal can be framed within the context of boundary methods.

\medskip

The article is organized as follows. Section 2 provides a detailed description of the glass data. In Section 3 a new procedure for one-class classification is introduced and tested in a simulation study.
In Section 4, the proposed methodology is applied to the motivating example dataset. A final discussion on the obtained results concludes the paper.

\section{Glass data}

The glass dataset used in this paper comes from UCI repository and contains $n=138$ glass fragments, whereof 51 containers/tableware/headlamps (\emph{non-window}) and 87 \emph{window} (car and building) samples. Since all these observations derive from a crime scene and no fragments from potential offenders are recorded, we decide to use the \emph{window} set as the target class. In other words, we derive the one-class classification rule on window objects only and we consider the \emph{non-window} ones to evaluate the rule performances.
These fragments are characterised by $p=9$ features: the Refractive Index and the chemical composition of 8 crucial elements, sodium ($Na$), magnesium ($Mg$), aluminium ($Al$), silicon ($Si$), potassium ($K$), calcium ($Ca$), barium ($Ba$) and iron ($Fe$). Each element is normalised to oxygen ($O$) so as to remove any stochastic fluctuation in instrumental measurements. Such features exhibit a moderately high correlation, as shown in Table \ref{tab:corr}.

\begin{table} \caption{Glass data: correlation matrix} \label{tab:corr}
\centering
\fbox{
\begin{tabular}{rrrrrrrrrr}
 & $RI$ & $Na'$ & $Mg'$ & $Al'$ & $Si'$ & $K'$ & $Ca'$ & $Ba'$ & $Fe'$ \\
  \hline
  $RI$ & 1.000 & 0.565 & 0.433 & -0.697 & -0.772 & -0.781 & 0.842 & 0.063 & -0.046 \\
  $Na'$ & 0.565 & 1.000 & 0.402 & -0.574 & -0.790 & -0.711 & 0.369 & 0.135 & -0.193 \\
  $Mg'$ & 0.433 & 0.402 & 1.000 & -0.437 & -0.484 & -0.540 & 0.186 & 0.007 & -0.130 \\
  $Al'$ & -0.697 & -0.574 & -0.437 & 1.000 & 0.506 & 0.770 & -0.703 & 0.032 & 0.041 \\
  $Si'$ & -0.772 & -0.790 & -0.484 & 0.506 & 1.000 & 0.720 & -0.673 & -0.170 & 0.078 \\
  $K'$ & -0.781 & -0.711 & -0.540 & 0.770 & 0.720 & 1.000 & -0.706 & -0.167 & 0.078 \\
  $Ca'$ & 0.842 & 0.369 & 0.186 & -0.703 & -0.673 & -0.706 & 1.000 & -0.026 & 0.039 \\
  $Ba'$ & 0.063 & 0.135 & 0.007 & 0.032 & -0.170 & -0.167 & -0.026 & 1.000 & -0.006 \\
  $Fe'$ & -0.046 & -0.193 & -0.130 & 0.041 & 0.078 & 0.078 & 0.039 & -0.006 & 1.000 \\
\end{tabular}}
\end{table}

\begin{figure}
\centering
\makebox{\includegraphics[width=.8\textwidth]{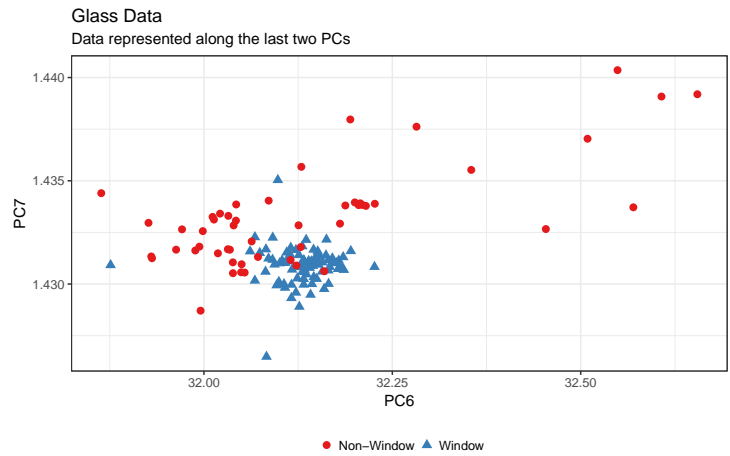}}
\caption{\label{fig:glassdata} Glass dataset. Data are projected on the last two principal components.}
\end{figure}

In order to evaluate how different the non-window are from the window samples, in Figure \ref{fig:glassdata} we plot the data according to the directions with the lowest variability, i.e. according to the last two principal components computed on the target set; this representation shows that the target class (the triangles) is quite compact, while samples from the outlier one (the circles) are scattered all around.

\begin{figure}
	\centering
	\makebox{\includegraphics[width=.8\textwidth]{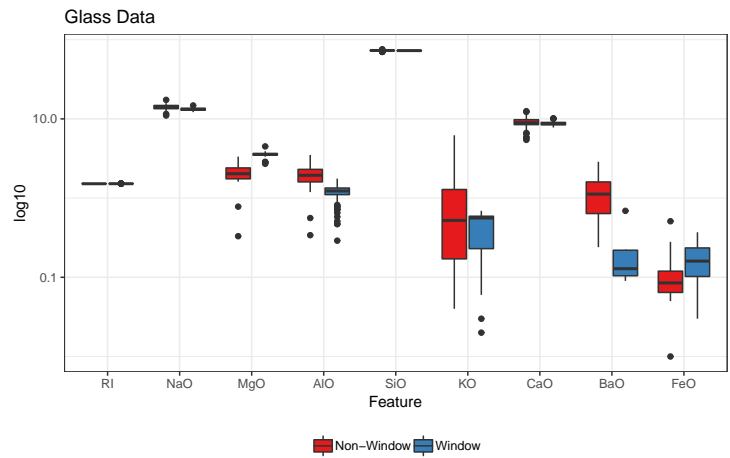}}
	\caption{\label{fig:boxplots} Feature distribution according to the sample type.}
\end{figure}

Figure \ref{fig:boxplots} shows the distributions of the features according to sample type; the variable-wise boxplots do not largely overlap, except for the $RI$ and the presence of silicon. Outlying samples exhibit overall a larger variability compared to the inlying ones.

\section{The proposal}
As discussed in the previous section, the goal of any one-class classifier is to define a classification rule that accepts as many \emph{target} objects as possible and rejects all those significantly deviating from this class. The crucial aspect that should be stressed is that one-class algorithms learn the classification rule by using a training set composed of a single class of well-known observations that does not include any anomaly.
Therefore, this issue is substantially different from a traditional two-class classification problem, where the aim is to assign data objects to one of two preliminarily defined categories.
It also differs from an outlier detection task, where the training set is naturally polluted by deviant observations.

In this work, a new statistical approach for one-class classification based on Gini's definition of \emph{transvariation probability} between a group and a constant is proposed. In particular, we refer to the concept of \emph{transvariation} and to some of its related measures, firstly introduced in a univariate context by \cite{gini1916concetto} and, subsequently, extended to the multivariate case and to a model-based formulation by \cite{gini1943transvariazione} and \cite{dagum1959transvariazione}, respectively.

\subsection{Transvariation probability as a measure of resemblance}

The transvariation concept has proved to be very useful in the standard classification context as a measure of group separability, especially when the assumptions that justify the optimality of Fisher's linear discriminant function are not met \citep{montanari2004linear}.
Its applicability can be even extended to the one-class domain, as the definition of transvariation probability seems to perfectly fit the idea of resemblance between an object and a group.
Moreover, this concept can be also viewed as a \emph{data depth} measure, i.e. a measure of how deeply a generic observation lies in the data cloud \citep{tukey1975}.

According to Gini~\citep{gini1916concetto},

\begin{definition}
	\label{definition}
	A group ${g}$ of $n$ units and a constant $c$ are said to
	transvariate on a variable $X$, with respect to a generic mean value ${m_X}$ if the sign of some of the ${n}$
	differences ${x_i-c}$, $i=1, \cdots, n$, is opposite to that of ${m_X-c}$.
\end{definition}
In this definition, the constant $c$ can be seen as the observed value of a {\em degenerate} group, that is, a group made of a single unit. Rephrasing such definition in the one-class domain, $c$ becomes the single unit whose resemblance with respect to the target class (namely, with $m_X$) shall be evaluated.

\begin{figure}
	\centering
	\makebox{\includegraphics[width=0.8\textwidth]{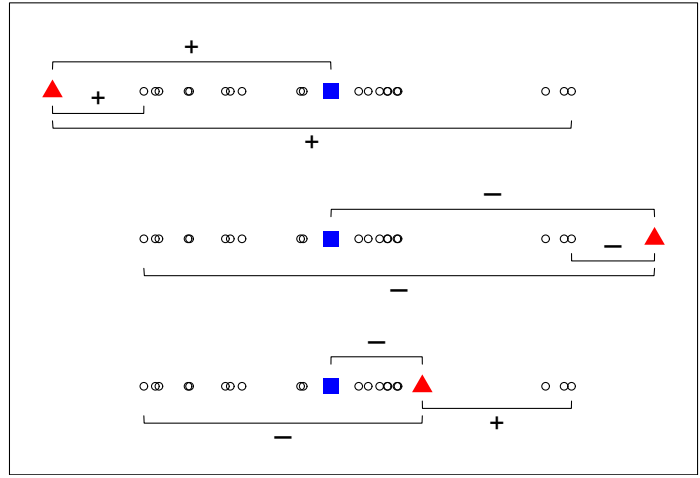}}
	\caption{\label{fig:transvariation} Two examples of no transvariation (first two rows) and a case of
transvariation (third row) between a given unit (the triangle) and the group median (the square).}
\end{figure}

In order to fully understand what transvariation means, consider as an example, the three different scenarios depicted in Figure~\ref{fig:transvariation}. In the first two, no transvariation occurs between constant $c$ (the triangle) and the mean value $m_X$ (the square) as all the differences $x_i-c$ (where $x_i$ is any group observation) have the same sign pattern. In the third case, on the contrary, there is evidence of transvariation: there are three points on the right-hand side whose differences with $c$ have opposite sign with respect to that of $m_X-c$.

\medskip
The probability that an event fulfills Definition~\ref{definition} is known as {\em transvariability}, $\tau$.
$\tau$ is simply the number of transvariations over the number of possible differences,
\begin{equation}
\label{trasnvariabilityd}
\tau = \frac{s_{X} + \frac{s'_{X}}{2}}{n},
\end{equation}

where: \begin{itemize}
	\item[-] $s_{X}$ is the number of units for which $(x_i - c)(m_X-c) < 0$;
	\item[-] $s'_{X}$ is the number of units for which $(x_i - c)(m_X-c) = 0$;
	\item[-] $n$ is the number of differences $(x_i - c)$.
\end{itemize}

If we assume $m_X$ to be the median (as Gini did), the maximum of $\tau$, $\tau_M$, is $\frac{1}{2}$.
Therefore, the definition of transvariation probability of a constant $c$, $tp(c)$, with respect to a group represented by its median is:

\begin{equation}\label{tp_discreta}
  tp(c) =  \frac{\tau}{(1/2)} = 2 \hspace{0.1cm} \frac{s_{X} + \frac{s'_{X}}{2}}{n} .
\end{equation}
Values close to 1 reflect a high resemblance of $c$ to the target class.

When the probability density function of the target class is known or can be estimated, an analogous version of transvariability ($\tau_f$) that exploits such information can be derived:

\begin{equation} \label{transvariabilitydb}
\tau_f = \min [F(c), 1 - F(c)] ,
\end{equation}
where $F(c)$ is the cumulative distribution function of the target class evaluated in $c$. Assuming $m_X$ to be the median, its maximum is still $\frac{1}{2}$ .
The resulting computation of transvariation probability is:
\begin{equation}
\label{tp_density}
tp_f(c) = \frac{\tau_f}{(1/2)} = 2 \cdot \begin{cases}F(c) &  m_X \ge c \\ 1 -F(c) &  m_X<c \end{cases}.
\end{equation}

\subsubsection{Extension to the multivariate case}

Transvariation probability allows for extensions to more than one variable. Specifically, following \cite{gini1943transvariazione}, in the multivariate case, the definition of transvariability $\tau$ corresponds to the {\em joint} probability that an event fulfills Definition~\ref{definition}:

\begin{equation} \label{taumult} \tau = \frac{s_{\mathbf X} + \frac{s'_{\mathbf X}}{2}}{n}, \end{equation}
where \begin{itemize}
	\item[-] $s_{\mathbf X}$ is the number of units for which $({x}_{iu} - {c_u})({m}_{u}-{c_u}) < 0$ for all the variables $u=1,\ldots,p$;
	\item[-] $s'_{\mathbf X}$ is the number of units for which $({x}_{iu} - {c_u})({m}_{u}-{c_u}) = 0$ for all the variables $u=1,\ldots,p$;
	\item[-] $n$ is the number of differences $({x}_{iu} - {c_u})$.
\end{itemize}

If we assume
\[ {\mathbf m}_{\mathbf X}=(m_1, \dots, m_p)\]
to be the multivariate {\em spatial} median or {\em mediancentre} \citep{mediancentre}, i.e. ${\mathbf m}_{\mathbf X}$ is the vector that minimizes $\sum_n d({\mathbf x},{\mathbf m}_{\mathbf X})$, where $d({\mathbf x},{\mathbf m}_{\mathbf X})$ is the distance between ${\mathbf x}$ and ${\mathbf m}_{\mathbf X}$, the maximum $\tau_M$ may no longer be $\frac{1}{2}$ and it needs to be estimated.
In particular, $\tau_M$ can be computed as $\tau$ in equation \ref{taumult} on the shifted data ${\mathbf Y} = {\mathbf X} - ({\mathbf m}_{\mathbf X}-{\mathbf c})$.
Therefore, the \emph{multivariate} definition of transvariation probability is:
\begin{equation} \label{tpmultd}  tp(\mathbf c) =  \frac{s_{\mathbf X} + \frac{s'_{\mathbf X}}{2}}{s_{{\mathbf Y}} + \frac{s'_{{\mathbf Y}}}{2}}. \end{equation}

Equation \ref{transvariabilitydb} can be extended to the multidimensional case as well. Given that $\tau_M$ may no longer be $\frac{1}{2}$, the expression of~(\ref{tp_density}) becomes:
\begin{equation}
\label{tpmultdb}
tp_f(\mathbf c)= \frac{\int_{a_{{\mathbf x}_1}\phantom{M}}^{b_{{\mathbf x}_1}\phantom{M}} \cdots  \int_{a_{{\mathbf x}_p}\phantom{M}}^{b_{{\mathbf x}_p}\phantom{M}}  f({\mathbf x}) \, d{\mathbf x}}{\int_{a_{M{\mathbf x}_1}}^{b_{M{\mathbf x}_1}} \cdots  \int_{a_{M{\mathbf x}_p}}^{b_{M{\mathbf x}_p}} f({\mathbf x}) \, d{\mathbf x}}
\end{equation}
where $f({\mathbf x})$ is the probability density function (pdf) of the target class and
\begin{multicols}{2}
	\begin{itemize}
	\item[-] $a_{{\mathbf x}_u\phantom{M}} = \begin{cases} c_u & \mbox{if } c_u \ge m_u \\ -\infty  & \mbox{if } c_u < m_u \end{cases}$,
	\item[-] $b_{{\mathbf x}_u\phantom{M}} = \begin{cases} +\infty & \mbox{if } c_u \ge m_u \\ c_u  & \mbox{if } c_u < m_u \end{cases}$,
	\item[-] $a_{M{\mathbf x}_u} = \begin{cases} m_u & \mbox{if } c_u \ge m_u \\ -\infty  & \mbox{if } c_u < m_u \end{cases}$,
	\item[-] $b_{M{\mathbf x}_u} = \begin{cases} +\infty & \mbox{if } c_u \ge m_u \\ m_u  & \mbox{if } c_u < m_u \end{cases}$,
\end{itemize}
\end{multicols}
for $u = 1,\dots,p$. Obviously, when the variables involved in the computation can be assumed to be independent, the multivariate transvariation probability reduces to the product of the simple univariate ones:
\[ tp(\mathbf c) = \prod_{u} tp(c_u) \quad u = 1, \dots, p,\]
where $tp(c_u)$ is the \emph{univariate} marginal transvariation probability corresponding to the $u$-th variable, computed either by (\ref{tp_discreta}) or (\ref{tp_density}).

\subsection{Transvariation-based One-Class Classifier (TOCC)}
In this paper, a new one-class classification method based on transvariation probability, called \emph{Transvariation-based One-Class Classifier} (TOCC), is introduced.
In particular, we shall refer to TOCC$_{df}$ if the transvariation probability is computed according to (\ref{tpmultd}) and thus it is \emph{density-free}; coherently, we would refer to TOCC$_{db}$ when considering equation~(\ref{tpmultdb}), as it is \emph{density-based}.

The classification rule of the TOCC$_{df}$ [TOCC$_{db}$] is obtained through the following steps:

\begin{enumerate}
	\item Set a value, $s$, as the desired minimum sensitivity of the one-class classifier;
	\item For each unit ${\mathbf c}$ compute its transvariation probability $tp(\mathbf c)$ [$tp_f(\mathbf c)$] with respect to the target group median, ${\mathbf m}_{\mathbf X}$;
	\item Use the $s-th$ percentile of the distribution of transvariation probabilities as a threshold, ${t}$, for the one-class classifier.
\end{enumerate}

For a new test sample ${\mathbf z}$, its transvariation probability, $tp(\mathbf z)$ [$tp_f(\mathbf z)$], with respect to $\mathbf{m}_{\mathbf X}$ is computed. Then, ${\mathbf z}$ is assigned to the target set if

\begin{equation}\label{eq:decision}
   tp(\mathbf z) \ge {t} \qquad \qquad [tp_f(\mathbf z) \ge {t}].
\end{equation}

\begin{figure}
	\centering
	\makebox{\includegraphics[width=.8\textwidth]{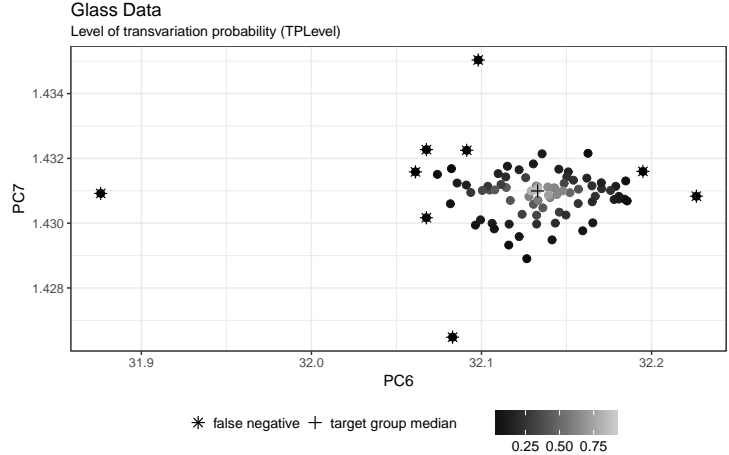}}
	\caption{\label{fig:transvariationlevel}Level of transvariation probability between each target observation and the target group median (the cross). Stars represent the objects (about 10\% of the whole target set) that are labelled as non-target.}
\end{figure}

In order to visualize how the TOCCs work in practice, consider Figure~\ref{fig:transvariationlevel}. In the plot, target glass samples are colored in different shades of gray, according to the level of their transvariation probabilities, $tp(\mathbf c)$, with respect to the target group median, ${\mathbf m}_{\mathbf X}$ (the cross). As expected, moving away from ${\mathbf m}_{\mathbf X}$, the magnitude of transvariation probability decreases. In particular, by setting $s=0.90$, all the objects with a value of $tp(\mathbf c)$ smaller than the threshold ${\mathbf t}$, are classified as (false) negative (i.e. the stars).

\bigskip

Consider again Figure~\ref{fig:glassdata}. As it can be easily noticed, the triangle cloud (i.e. the target class) is polluted by several non-target objects. As the TOCC can be seen as a data depth measure, it tends to `peel' the target set and, therefore, it may fail to detect those deviating observations that do not lie on the external border.
In order to improve this procedure, and inspired by those algorithms that use a \emph{set} of prototypes to represent the input data (e.g. $K$-means, SOM, \dots), a modified version of the TOCC$_{df}$ is introduced.

\begin{figure}
	\centering
	\makebox{\includegraphics[width=\textwidth]{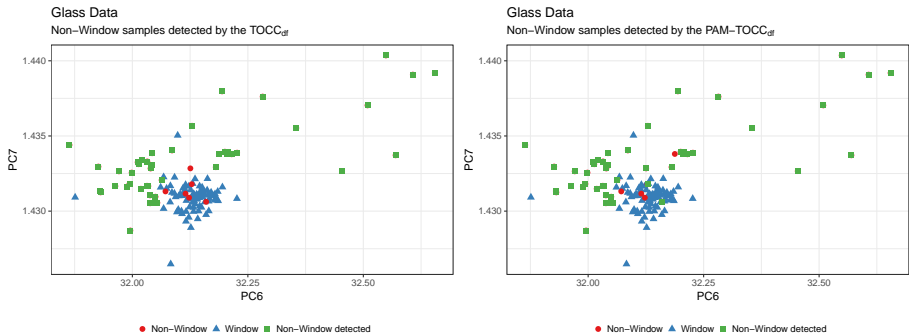}}
	\caption{\label{fig:outlierdetected}Class membership of the glass data predicted by the TOCC$_{df}$ (left panel) and the PAM-TOCC$_{df}$ (right panel) with a number of clusters $K=4$.}
\end{figure}

The idea is to combine the TOCC$_{df}$ with the clustering information on the target class provided by Partitioning Around Medoids, PAM \citep{kaufman1990}. Each cluster is analysed separately; as a result, the PAM-TOCC$_{df}$ returns a \emph{set} of thresholds, rather than a single one. In so doing, the algorithm is capable to detect those deviating observations that are scattered within the target set.

Figure~\ref{fig:outlierdetected} shows the two different solutions yielded by the the TOCC$_{df}$ and the PAM-TOCC$_{df}$. As discussed, the TOCC$_{df}$ (left panel) is able to identify only those deviating points placed on the target class perimeter. For this reason, such procedure is suggested when there is no evidence of strong overlap between the two sets. In all the other situations, the PAM-TOCC$_{df}$ (right panel) should be preferred: as clearly displayed, this algorithm is able to detect non-target objects that deviate along different directions.

\medskip

The following steps outline the  PAM-TOCC$_{df}$ two-phases process:
\begin{enumerate}
	\item[] {\bf Phase I}:
	\begin{enumerate} \item run the PAM algorithm on the target class, with a number of clusters $K$ chosen beforehand; store the resulting information on both the group membership and the prototype vectors. \end{enumerate}
	\item[] {\bf Phase II}: for each cluster $k$,
	\begin{enumerate} \item set a value, $s_k$, as the desired minimum sensitivity of the one-class classifier (generally, $s_k$ is set equal $\forall k$);
		\item for each unit ${\mathbf c}$ in the $k$-th cluster compute its transvariation probability $tp(\mathbf c)$ with respect to the group prototype, $_k{\mathbf m}_{{\mathbf X}}$. As ${\mathbf m}_{\mathbf X}$ is no longer the median, but the cluster centroid, there is no guarantee that $\tau_M$ is equal to $\frac{1}{2}$. For this reason, the transvariation probability should be computed according to equation \ref{tpmultd}, in both the univariate and the multivariate contexts;
		\item use the $s_k-th$ percentile of the (increasing) ordered distribution of transvariation probabilities as a threshold, $_k{\mathbf t}$, for the one-class classifier. \end{enumerate}
\end{enumerate}

A new sample ${\mathbf z}$ is firstly assigned to the closest group $g$. Then, its transvariation probability, $tp(\mathbf z)$  with respect to $_g{\mathbf m}_{\mathbf X}$, is computed. The final decision on ${\mathbf z}$ is carried out according to the rule described in (\ref{eq:decision}), where ${t}= {_{k=g}{t}}$.

\subsection{Practical considerations} \label{sec:practcons}
The computational cost of the TOCCs increases with the number of features $p$ involved in the problem at hand.

For the TOCC$_{df}$ this relationship is (at most) \emph{linear}: the algorithm examines one variable at a time and, thus, it requires the calculation of (at most) $n\times p$ differences $({x}_{iu} - {c_u})({m}_{u}-{c_u}), \ i=1,\ldots,n$, \ $u=1,\ldots,p$, in order to decide whether the object $\mathbf{c}$ transvariates.

\begin{figure}
    \centering
    \begin{subfigure}[b]{0.45\textwidth}
        \includegraphics[width=\textwidth]{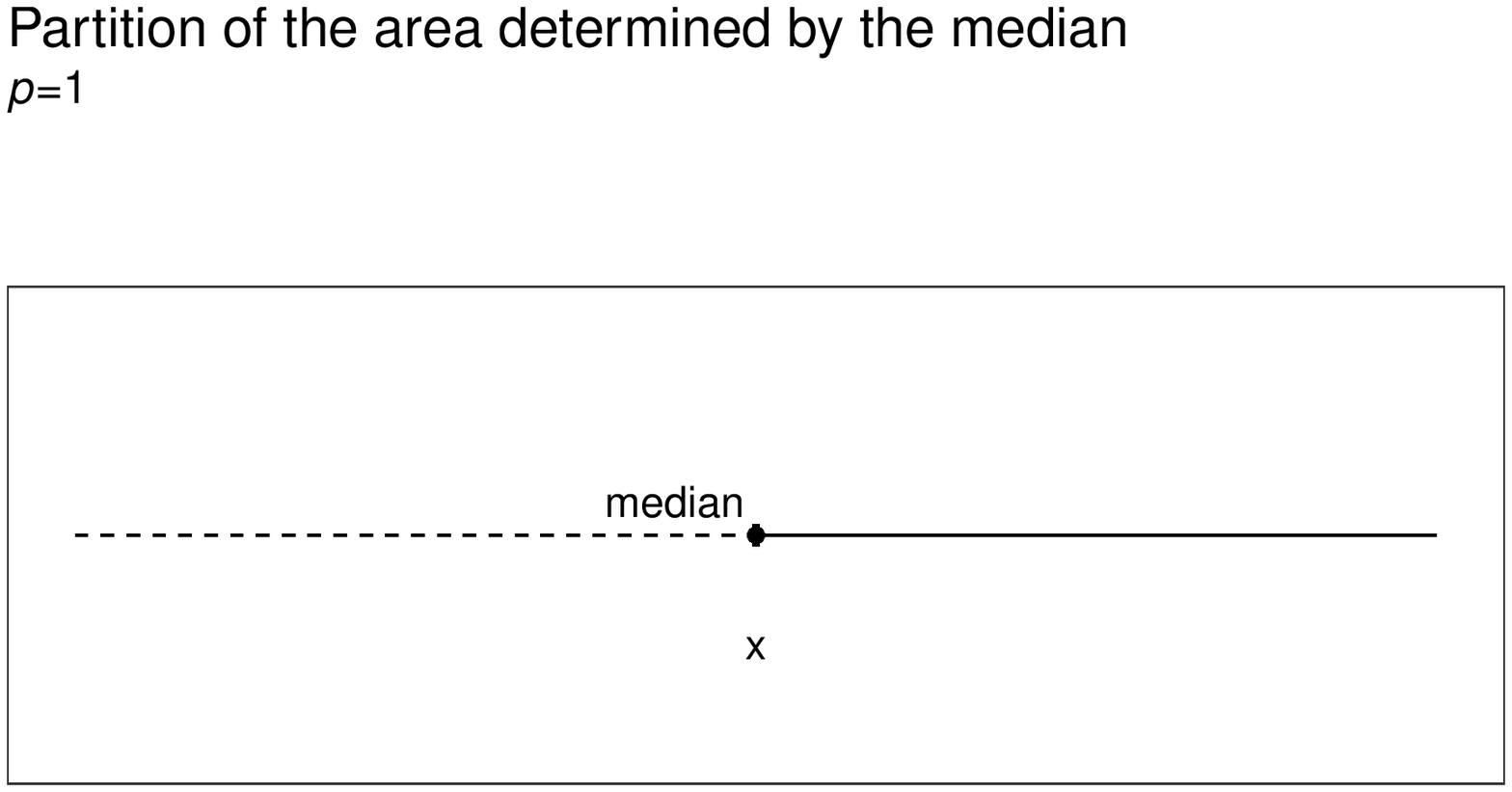}
        \label{fig:median_plot1}
    \end{subfigure}
    ~ 
    \begin{subfigure}[b]{0.45\textwidth}
        \includegraphics[width=\textwidth]{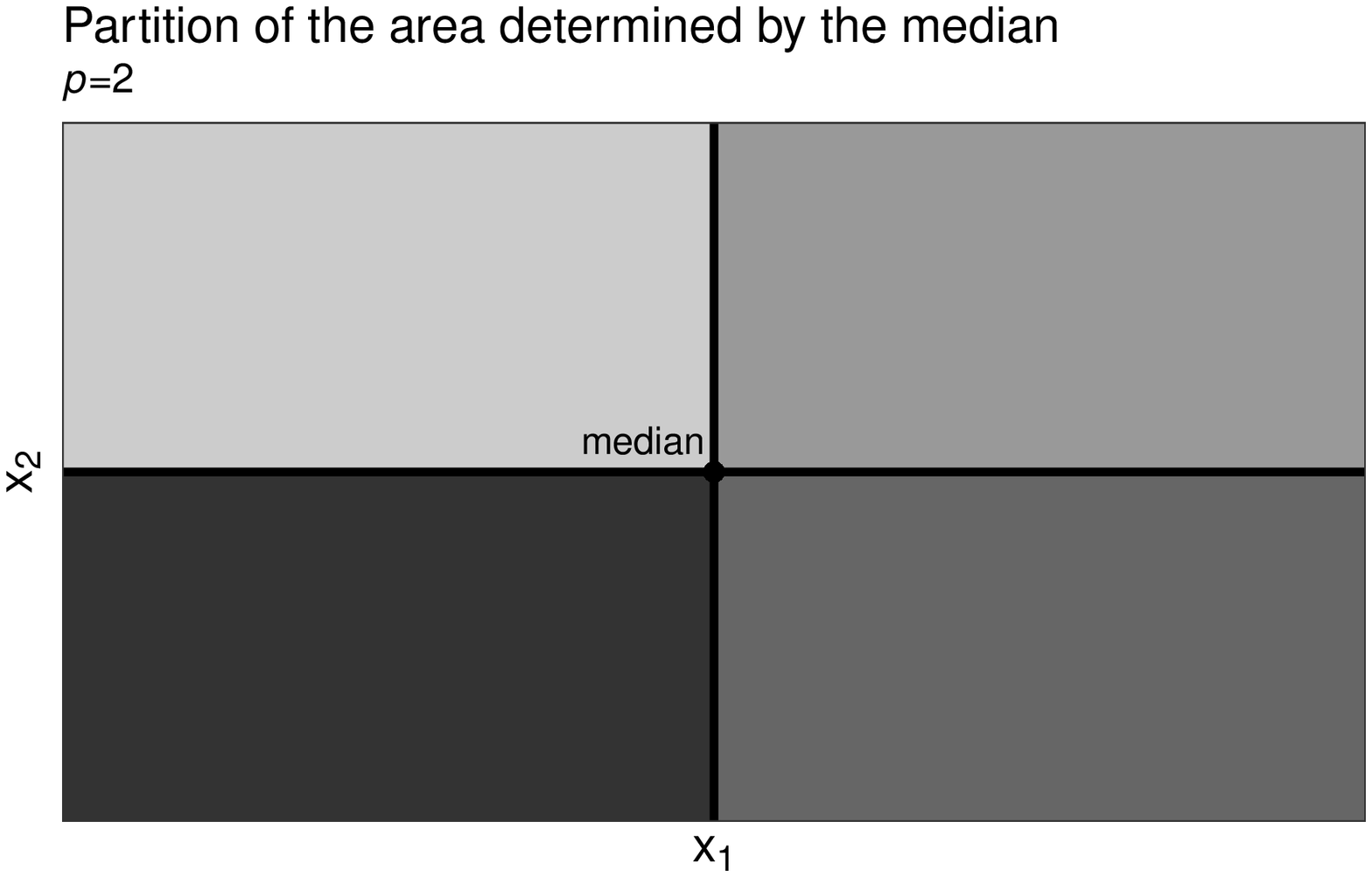}
        \label{fig:median_plot2}
    \end{subfigure}
    \caption{Representations of the total area split in $2^p$ regions by the median.}\label{fig:median_plots}
\end{figure}

In the case of the TOCC$_{db}$, the area under the curve is split into $2^p$ regions, identified at the intersection of the $p$ axes that originate from the spatial median, ${\mathbf m}_{\mathbf X}=(m_1, \dots, m_p)$, as shown in Figure \ref{fig:median_plots}.

Differently from the TOCC$_{df}$, the TOCC$_{db}$ is not a \emph{step-wise} procedure, as it considers all the variables together (see equation \ref{tpmultdb}). However, the cost of the algorithm increases \emph{exponentially} with $p$, since $2^p$ regions must be defined; unfortunately, this step is not scalable.

For these reasons, preliminary dimension reduction or variable selection procedures may be convenient in order to handle the classification task efficiently. In the following, several strategies are outlined.

\subsubsection{Dimension reduction and variable selection}
For dimension reduction, the classical Principal Component Analysis (PCA) or its sparse version (sPCA)  introduced by \cite{zou2006sparse} proved to produce good results in the one-class framework, especially when only the low-variance projections are retained \citep{tax2003feature}. In fact, such directions turned out to be the most informative ones for the one-class classification problem, since they provide the tightest description of the target set.

Besides PCA, the Random Projection (RP) method represents a valid alternative for reducing the data dimensionality. In the context of supervised classification, ~\cite{samworth} proposed an ensemble method that identifies the best $B_1$ RPs according to the smallest misclassification error rate. Within the one-class classification framework, a similar approach can be implemented. In this context the information on non-target objects is unavailable or vague, therefore a possible solution is to select those RPs that minimise the Median Absolute Deviation (MAD) of the projected data. Coherently with the definition of transvariation probability in (\ref{definition}), such strategy provides indeed the most compact projection of the target set with respect to its median. The resulting classification vectors are then aggregated through a majority vote scheme.

To deal with the variable selection task, many approaches have been developed in the model-based clustering and classification framework, e.g. \cite{scrucca2014clustvarsel}, \cite{murphy2010variable} and \cite{mclachlan2005analyzing}. Among them, \emph{varSel} algorithm introduced by~\cite{sartori} uses Gaussian Mixtures to identify the most suitable variables for classification (and clustering) purposes.

Random projections can also be exploited to perform variable selection. The input features could be ranked according to a modified version of the Importance Coefficient (CI) introduced by \cite{montanari2001projection} in the context of projection pursuit.  For the generic $d$-dimensional RP, the CI of the {\em u}-th variable is computed as:
\[
CI_{ui} = \sum_{q=1}^{d} \frac{|a_{uqi}|s_u}{\sqrt{\sum_{z=1}^{p}\left(a_{uzi}s_u\right)^2}}
\]
where $a_{uqi}$ indicates the attribute {\em u} coefficient in the {\em q}-th vector of the {\em d}-dimensional random projection solution $i$ and $s_u$ the variability (i.e. the standard deviation) of each attribute.
Since $B_1$ random projections are available, the overall importance measure for each variable can be derived as the median CI across projections and it is called \emph{Variable Importance in Projection} (VIP):

\begin{equation}
\label{VIP}
\mbox{VIP}_u = \underset{_{i=1,\dots,B_1}}{\mathrm{median}} \hspace{0.2cm} CI_{ui}.
\end{equation}
The median is used here so as to mitigate the effects of potential not-so-good projections on the VIP. The number of variables to be kept is decided by the user.

\medskip
The presence of highly associated input features pollutes the capability of the VIP to detect those actually relevant since, by its nature, it tends to assume approximately the same value for very correlated variables. Thus, a specific correction procedure for this measure is advisable in order to mitigate the correlation effect.

A possible strategy is to retain the variables with the highest VIP value whilst discarding those that strongly correlate, on average, with the variables already considered; i.e. those that exhibit an average absolute correlation $\bar{\rho}$ larger than a given threshold, $\kappa$. From our empirical experience, a reasonable interval for $\kappa$  would be $0.4-0.7$, depending on the average degree of the association in the original data: the strongest the association, the lower is the threshold. We shall refer to the \emph{adjusted-for-correlation} VIP as the $\kappa-$VIP.

\subsection{Simulated examples}
The performances of the TOCCs have been evaluated in an extensive simulation study. In each of the simulation settings described below, target ($\chi$) objects are generated according to different bivariate distributions, so as to visualise how the proposals work in practise.
Non-target data ($\Upsilon$) are considered to evaluate the performances of the classification rules learned on $\chi$ only.

For the first four scenarios, the mean vector of non-target data is obtained by shifting the mean vector of target objects. The magnitude of the shift is described by a non-centrality parameter, called $\lambda$;  different magnitudes (i.e. $\lambda = 1$, small shift;  $\lambda = 2$, large shift) are considered.

\begin{enumerate}
	\setlength\itemsep{1em}
	\item In the first scenario, we simulate target objects from a bivariate Gaussian distribution, whose components are standard normal random variables with a correlation equal to 0.35.
	\item Second scenario considers a skew target class, i.e. the squared bivariate Gaussian distribution of scenario (a) is used as generative model.
	\item Differently, in the third scenario, target data are generated by taking the the square root of the absolute bivariate Gaussian distribution in scenario (a).
    \item In scenario four, data are drawn from the logarithm of the bivariate Gaussian distribution in scenario (a).
\end{enumerate}

Further settings have been explored, i.e. scenarios (e)-(h), so as to evaluate the behaviour of the TOCCs in the presence of non-target objects uniformly scattered within a box over the target class. The size of the box
is determined by the target data itself; basically, the center of the box is the median of the features, and the sides are 3 times the interquartile range of each dimension. The same distributions of scenarios (a)-(d) are considered as target class.

An additional scenario (i) with non-standard data shape is also evaluated. Specifically, in this case, both target and non-target objects are generated according to a bivariate \emph{banana-shaped} distribution with different angle widths.

\medskip	
For each scenario, different sizes of the target class, $n_{T}$, are considered (i.e. 100, 200, 500); non-target class size, $n_{NT}$, is always taken to be $0.5 n_T$. For each setting, 100 repetitions are run and results are compared with several state-of-the-art one-class classifiers.

In particular, these methods include the Gaussian model (Gauss, implemented using the \texttt{mahalanobis} function), the Mixture of Gaussians approach (Mix-Gauss, implemented using the \texttt{mclust} package \citep[see][]{mclust}, where the optimal number of components, ranging from 1 to 9, was chosen so as to maximize the BIC), the kernel density estimate (KDE, implemented using the \texttt{ks} package with the normal kernel and the unconstrained plug-in bandwidth selector), the $K$-means algorithm (KM, implemented using the \texttt{kmeans} function with $K=5$ clusters), the 2-dimensional self organizing map (SOM, implemented using the \texttt{kohonen} package with a $5\times5$ grid and a learning rate $\alpha = (0.5, 0.3)$) and the support vector data description (SVDD, implemented using the \texttt{svdd} package, with a cost parameter for the positive examples $C=0.1$).

Mixtures of Gaussians are fitted to the data for the TOCC$_{db}$ in each scenario. The PAM-TOCC$_{df}$ has run with a number of clusters $K=5$, coherently with the settings of the competing methods.

\begin{figure}
    \centering
    \begin{subfigure}[b]{0.49\textwidth}     \centering
        \includegraphics[width=\textwidth]{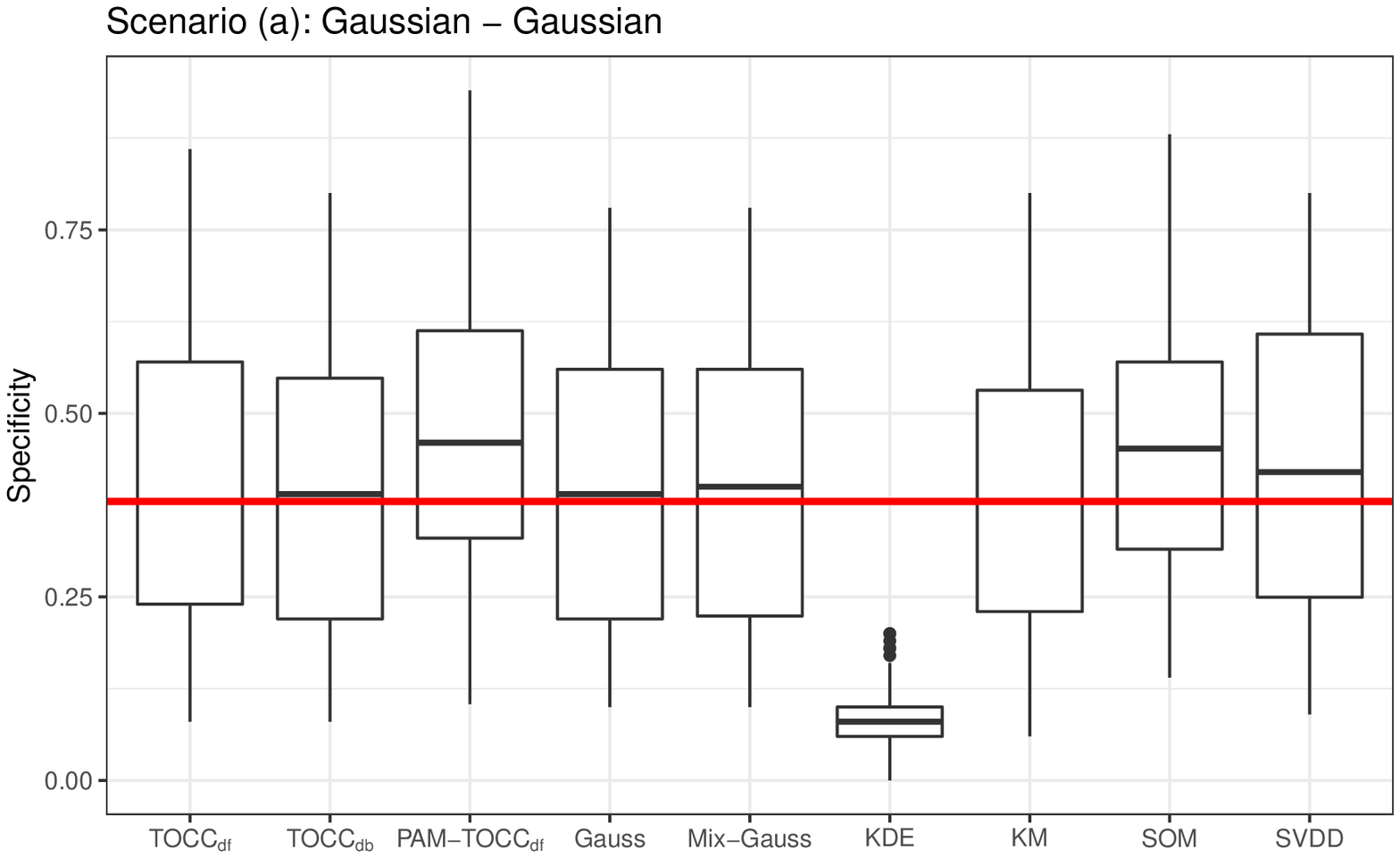}
        \label{fig:scenario1}
    \end{subfigure}
    \begin{subfigure}[b]{.49\textwidth}     \centering
        \includegraphics[width=\textwidth]{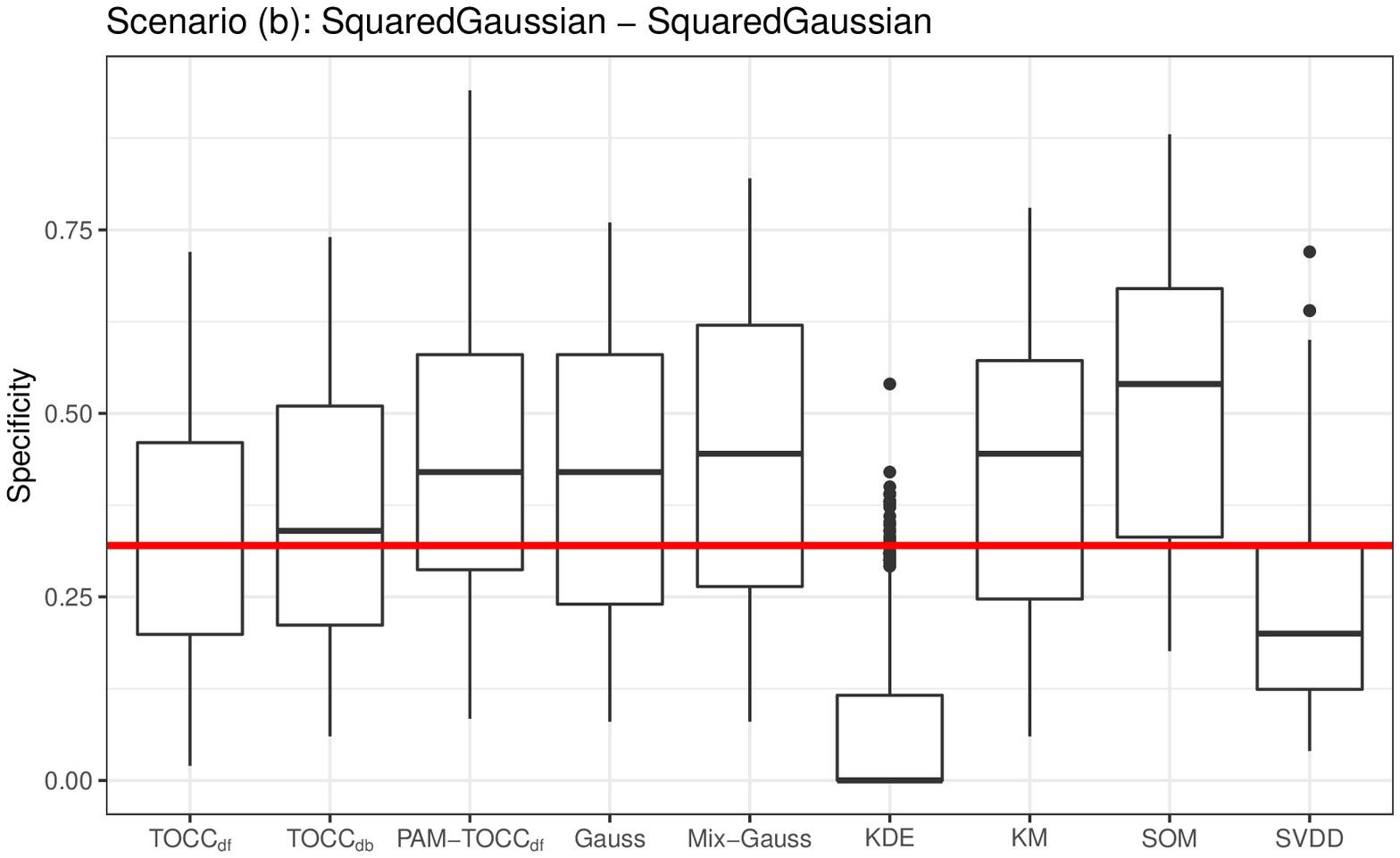}
        \label{fig:scenario2}
    \end{subfigure} \\
    \begin{subfigure}[b]{.49\textwidth}     \centering
        \includegraphics[width=\textwidth]{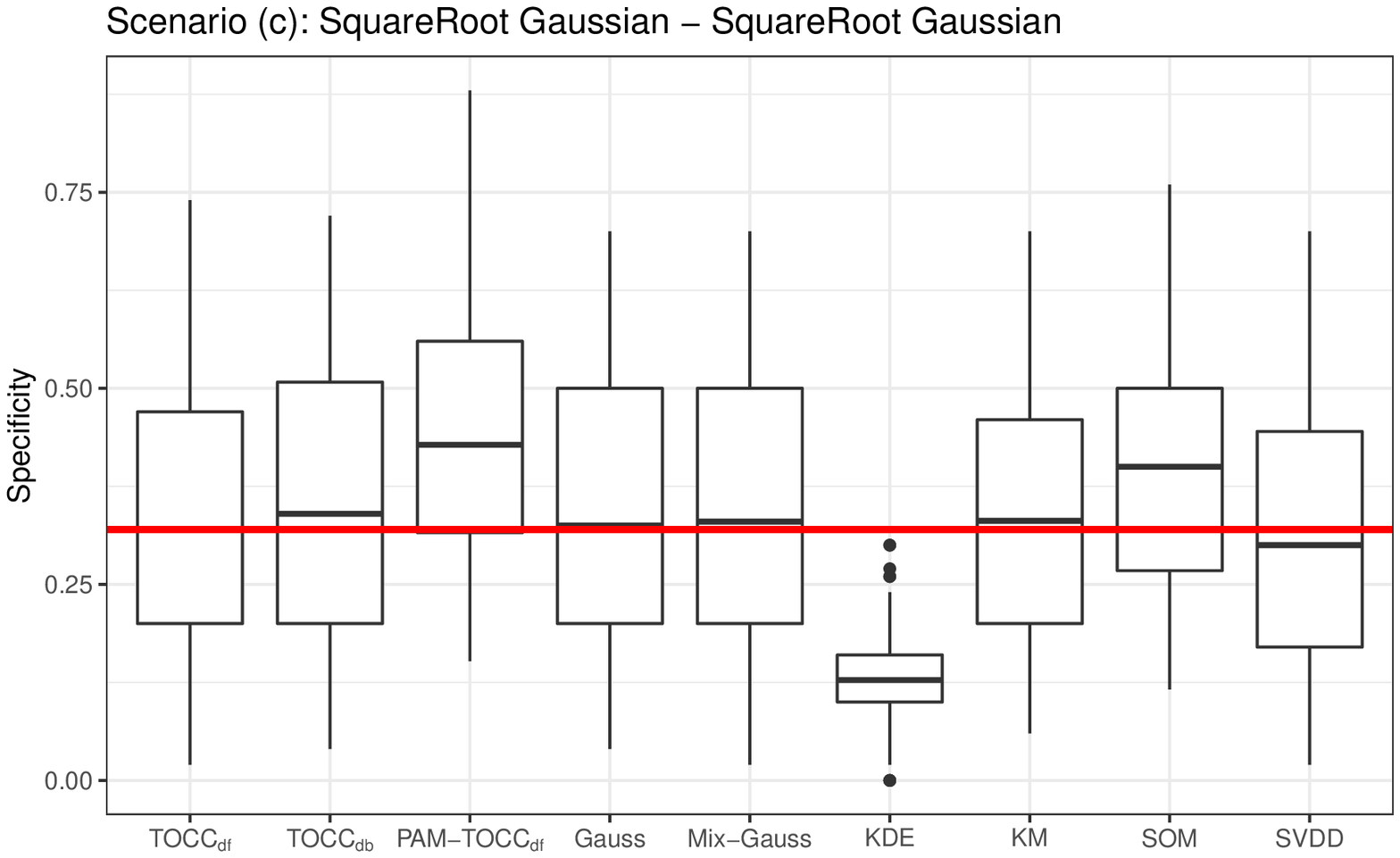}
        \label{fig:scenario3}
    \end{subfigure}
       \begin{subfigure}[b]{.49\textwidth}     \centering
        \includegraphics[width=\textwidth]{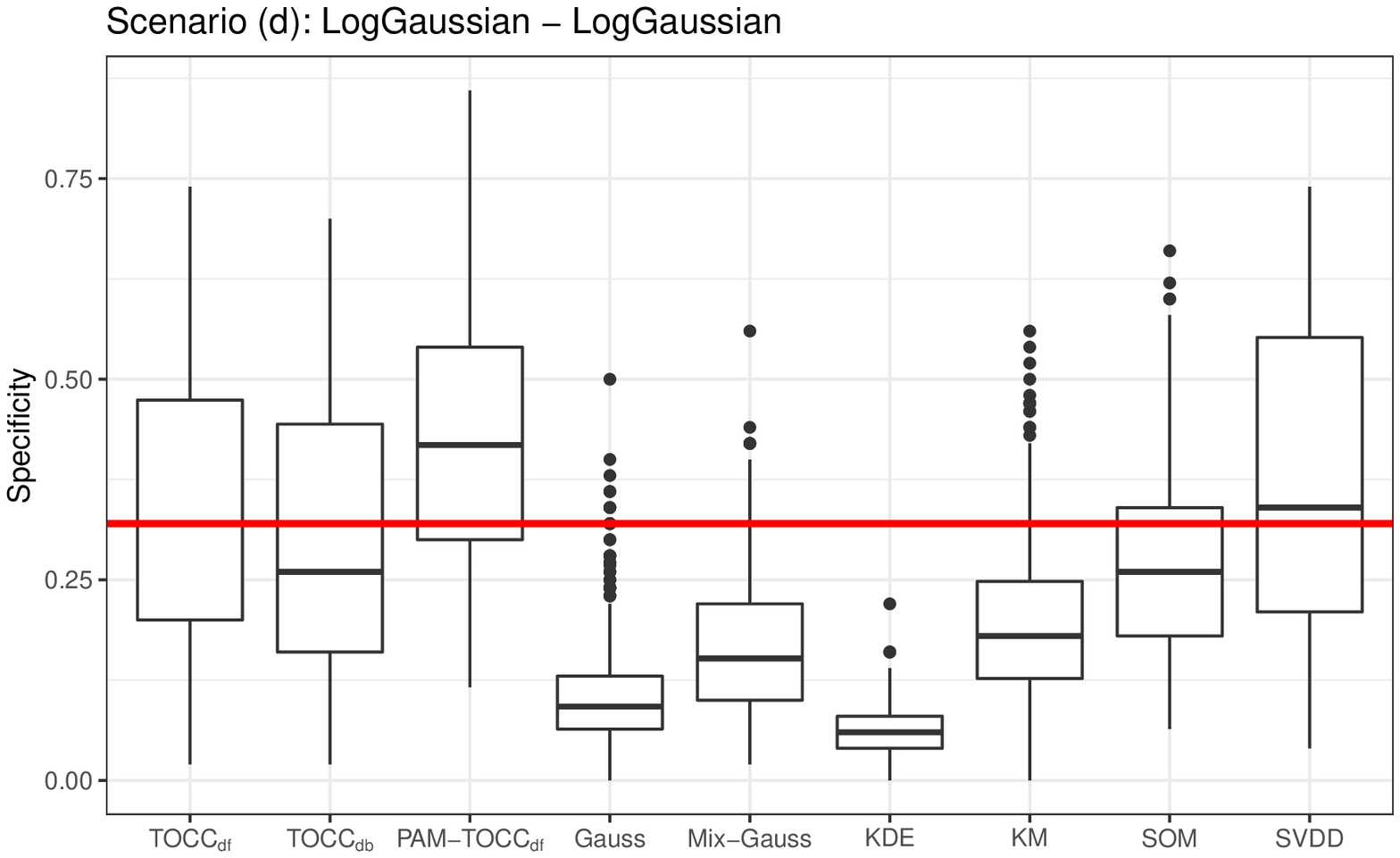}
        \label{fig:scenario4}
    \end{subfigure}
    \caption{Simulation results for scenarios (a) - (d): specificity rates for $s\geq 0.9$ sensitivity level. The horizontal line highlights the median specificity for the TOCC$_{df}$.}\label{fig:sim_plots1}
\end{figure}

\begin{figure}
    \centering
    \begin{subfigure}[b]{.49\textwidth}     \centering
        \includegraphics[width=\textwidth]{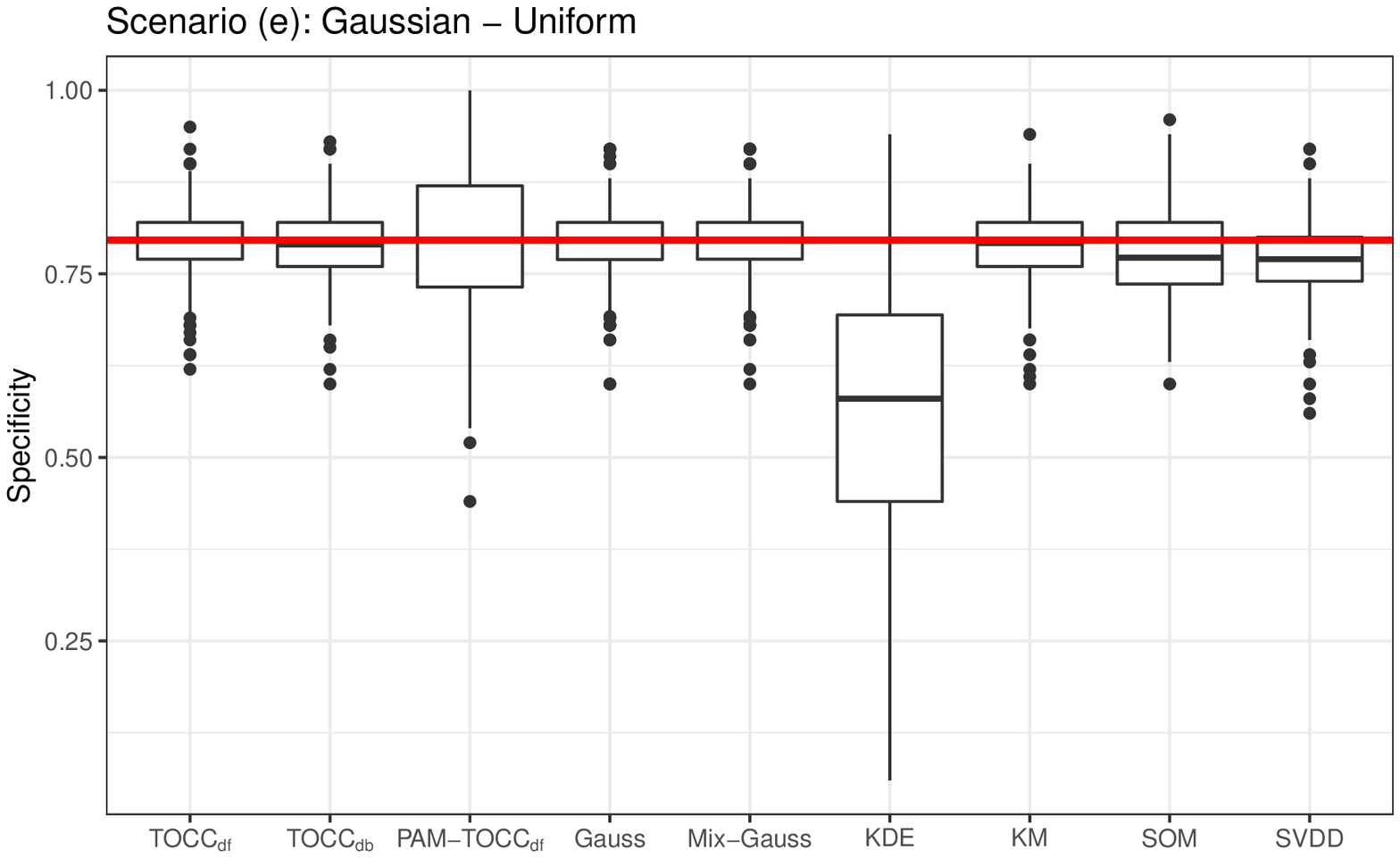}
        \label{fig:scenario5}
    \end{subfigure}
    \begin{subfigure}[b]{.49\textwidth}     \centering
        \includegraphics[width=\textwidth]{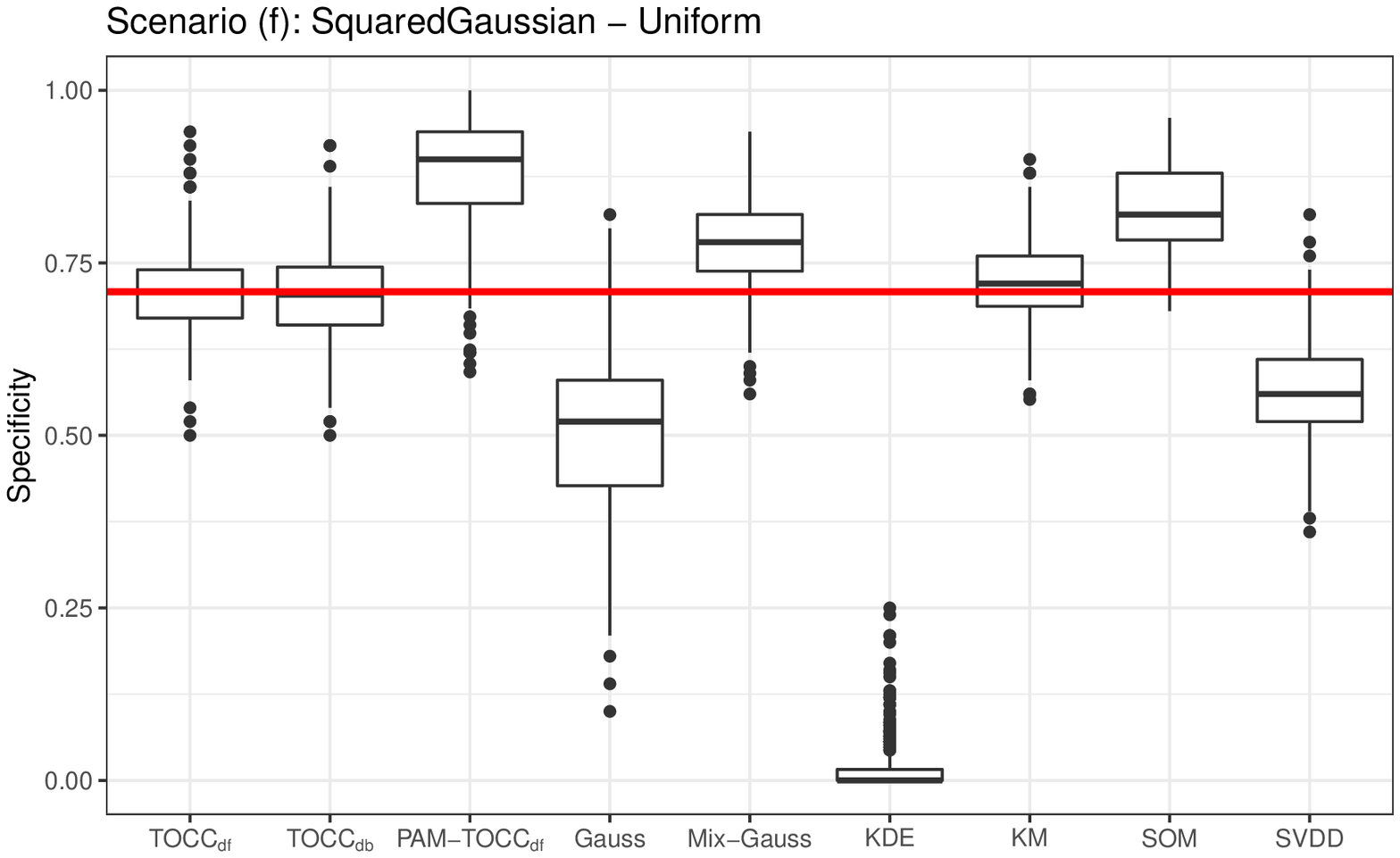}
        \label{fig:scenario6}
    \end{subfigure} \\
    \begin{subfigure}[b]{.49\textwidth}     \centering
        \includegraphics[width=\textwidth]{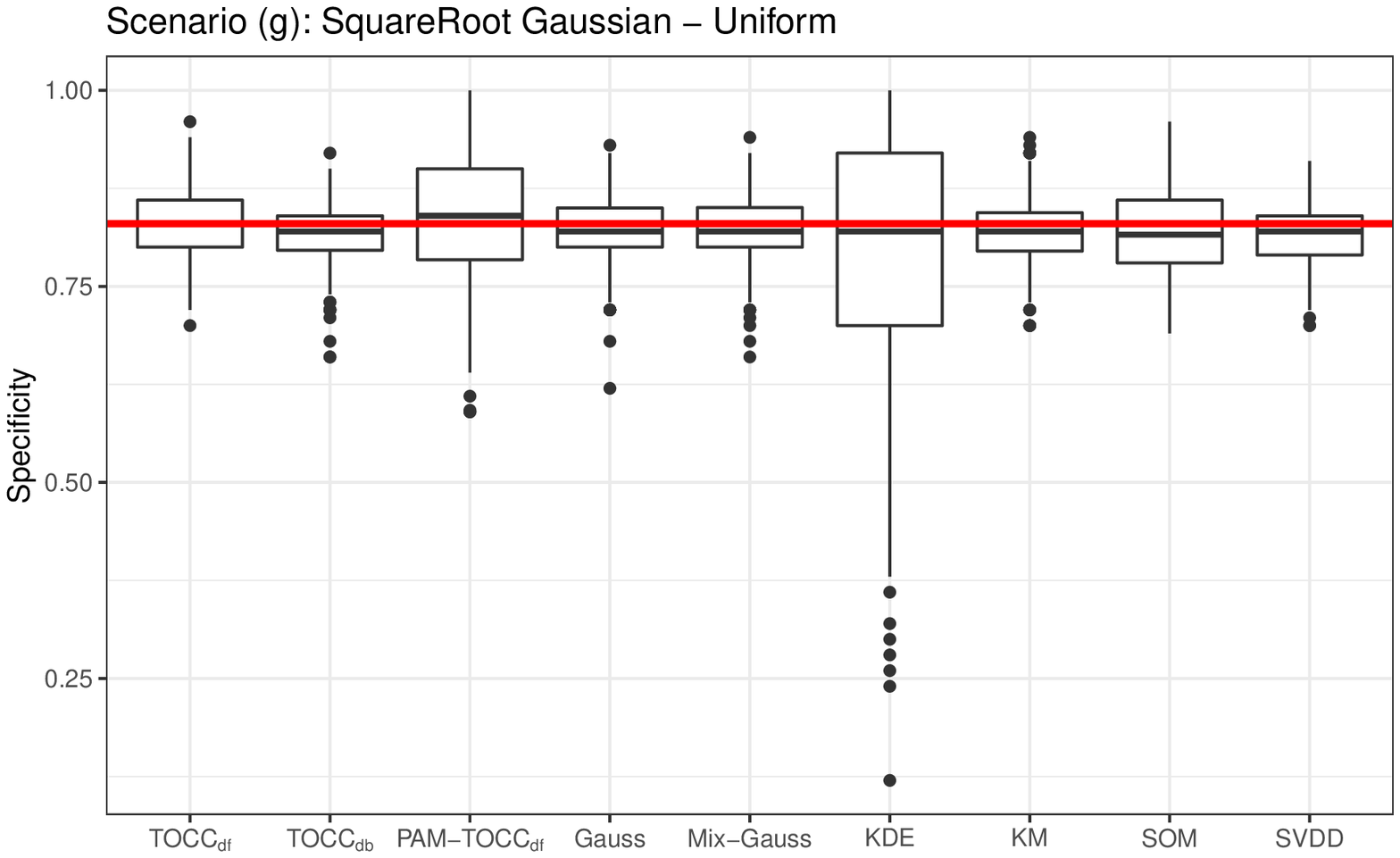}
        \label{fig:scenario7}
    \end{subfigure}
       \begin{subfigure}[b]{.49\textwidth}     \centering
        \includegraphics[width=\textwidth]{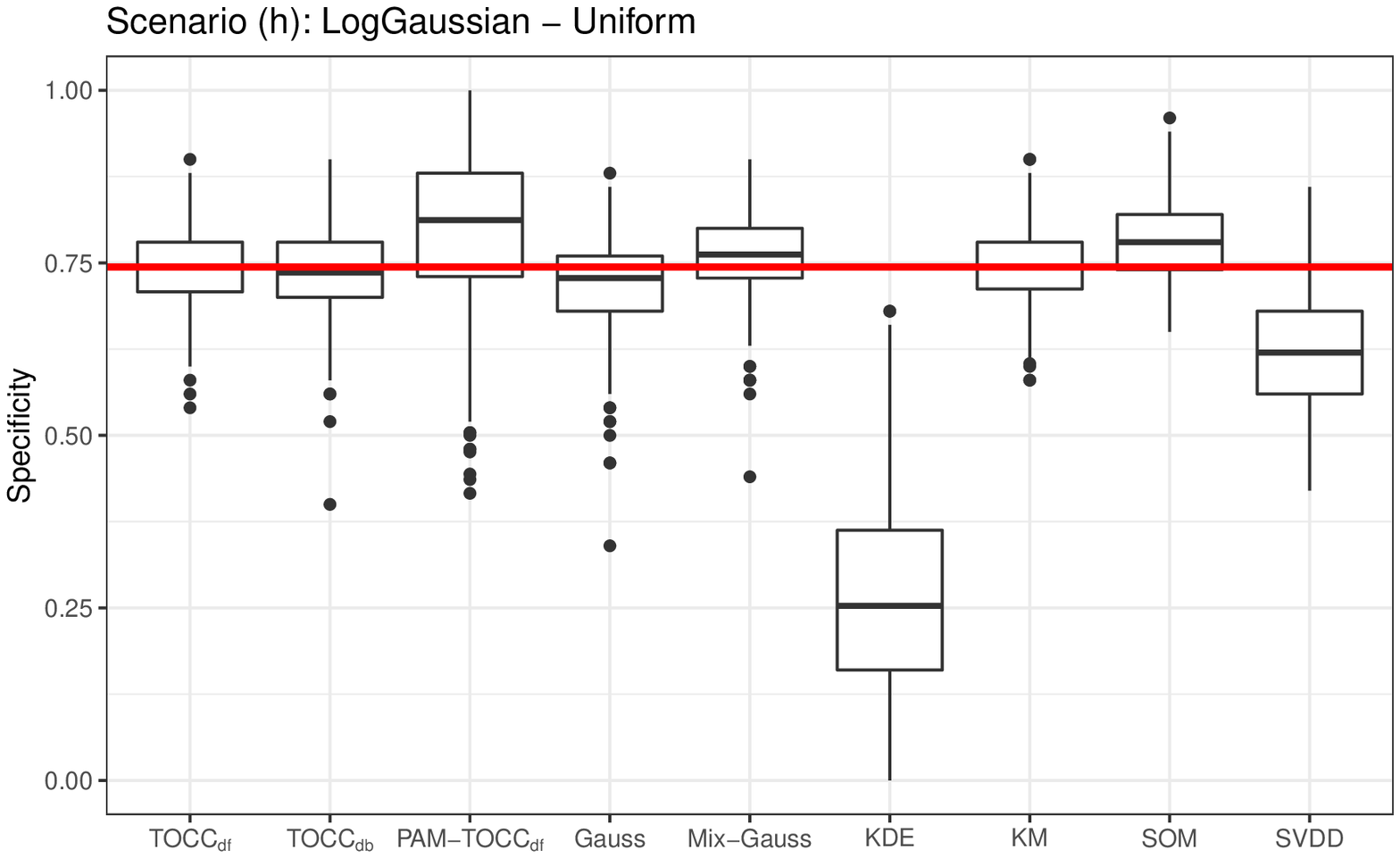}
        \label{fig:scenario8}
    \end{subfigure}
    \caption{Simulation results for scenarios (e) - (h): specificity rates for $s\geq 0.9$ sensitivity level. The horizontal line highlights the median specificity for the TOCC$_{df}$.}\label{fig:sim_plots2}
\end{figure}

\medskip
Figures \ref{fig:sim_plots1} and \ref{fig:sim_plots2} contain the aggregated results for each scenario. The boxplots show the behaviour of the specificity rates for $s\geq 0.9$ sensitivity level; the horizontal line helps the comparison among the approaches, by highlighting the median specificity for the TOCC$_{df}$.

Results coming from this study clearly show the general effectiveness of the transvariation-based one-class classifier we introduced. In particular, for all the simulated models, the algorithms attain specificity rates that are always better than or, at least, comparable with those from the state-of-the-art methods. These promising outcomes allow to efficiently use the proposed procedures in a wide variety of problems.

A separate evaluation should be carried out for the PAM-TOCC$_{df}$; the performances of this classifier strongly depend on the behavior of the non-target observations. In fact, as clearly depicted in the boxplots of Figure~\ref{fig:sim_plots1}, it tends to outperform the other methods especially when the detection problem is particularly difficult, that is, when non-target observations pollute the core of target set and do not limitedly lie on its external perimeter.

Boxplots in Figure \ref{fig:sim_plots2} exhibit a generally improved performance for almost all the methods in the presence of non-target samples uniformly scattered over the target set: overall, the median specificity for a sensitivity level $s\geq 0.9$ is above 75\%. Also in these scenarios, the PAM-TOCC$_{df}$ is able to globally detect the largest number of deviating observations.

Among the considered state-of-the art methods, the KDE appears to perform poorly almost everywhere. This is probably due to a wrong specification of the bandwidth matrix $H$ for the non-target class: $H$ is estimated only on the target set and, therefore, the kernel $\varphi_H(.)$ is likely to produce incorrect estimates for the observations that differ too much from this class.

\begin{figure}
    \centering
        \includegraphics[width=.6\textwidth]{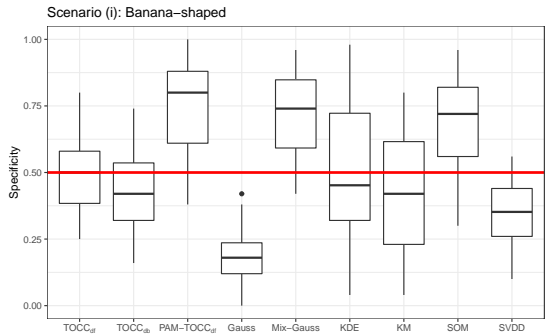}
         \caption{Simulation results for scenario (i): specificity rates for $s\geq 0.9$ sensitivity level. The horizontal line highlights the median specificity for the TOCC$_{df}$.}    \label{fig:sim_banana}

 \end{figure}

A special mention should be made for the results of the last scenario, depicted in Figure \ref{fig:sim_banana}. In general, the \emph{non-convexity} of the banana-shaped data appears very hard to be detected, particularly by the less flexible methods. In such situations, the most adaptive procedures (i.e. PAM-TOCC$_{df}$, Mix-Gauss and SOM) handle the ``non-typicality'' of the target class distribution more appropriately.

\section{Glass data analysis}
The analysis of the glass fragments is carried out by the TOCCs proposed and described in the previous sections.
Preliminarily, dimension reduction and variable selection procedures are applied, as suggested in Section \ref{sec:practcons}.

PCA is computed on the window fragments and the last two components are retained. For the RP method, the best $B_1 = 101$ bi-dimensional projections are considered, each carefully chosen within $B_2 = 50$ possible solutions.

About the variable selection procedures, the first two most important features according to both the \emph{VarSel} and the VIP algorithms are considered; in particular, given the moderately high degree of association (see Table \ref{tab:corr}), the \emph{adjusted-for-correlation} VIP is applied, with a threshold $\kappa=0.5$.

The bi-dimensional target data representation of Figure \ref{fig:glassdata} shows an approximately elliptical shape that suggests to consider a mixture of Gaussians as the reference model for this class.
As the chemical composition of the two sets of fragments is similar, we can expect them to be (at least) partially overlapping; for this reason, the PAM-TOCC$_{df}$ is run with a number of clusters moderately large compared to the number of units, i.e. $K=4$.

\begin{figure} \centering
	\begin{subfigure}[b]{.49\textwidth}     \centering
		\includegraphics[width=\textwidth]{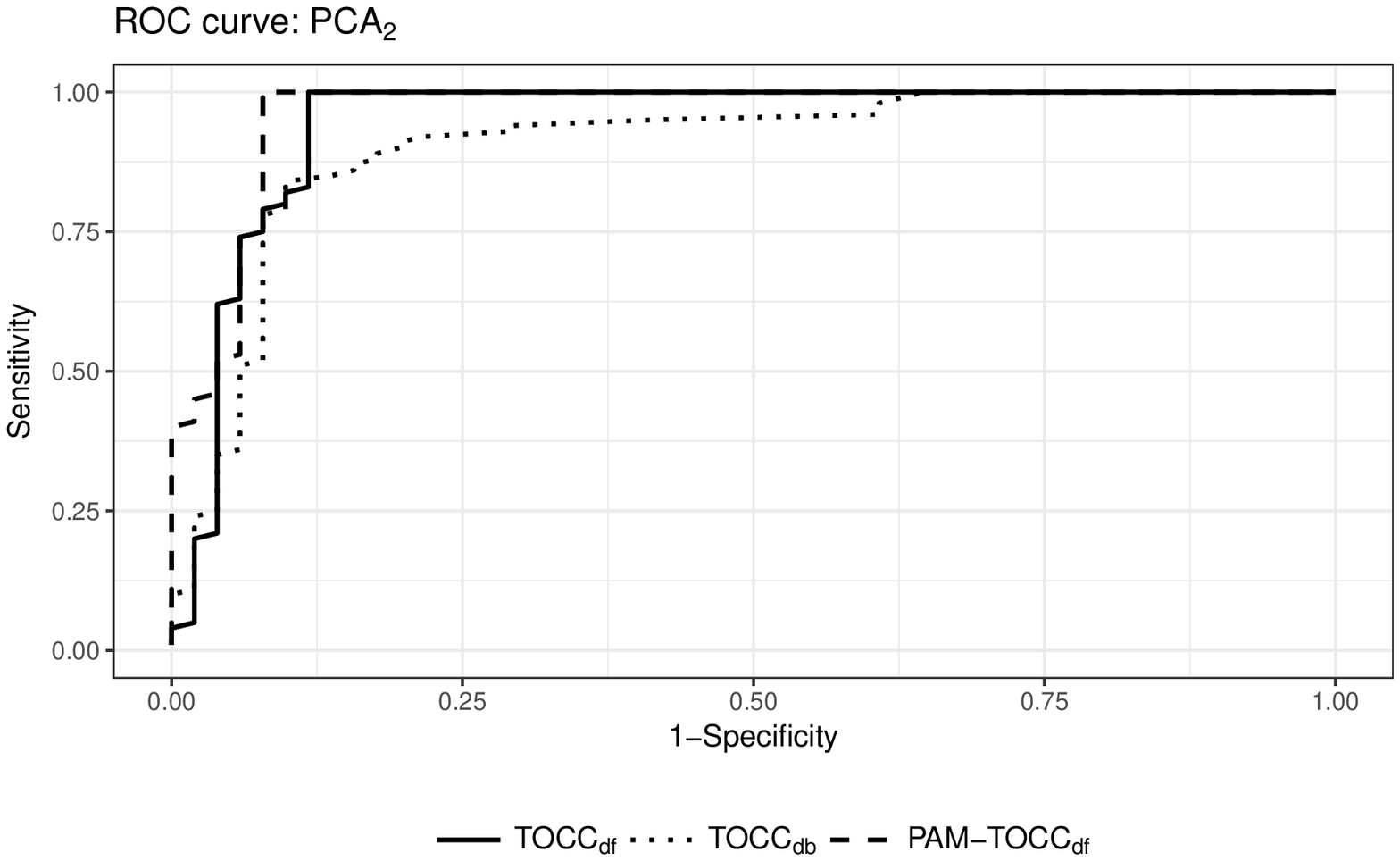}
	\end{subfigure}
	\begin{subfigure}[b]{.49\textwidth}     \centering
		\includegraphics[width=\textwidth]{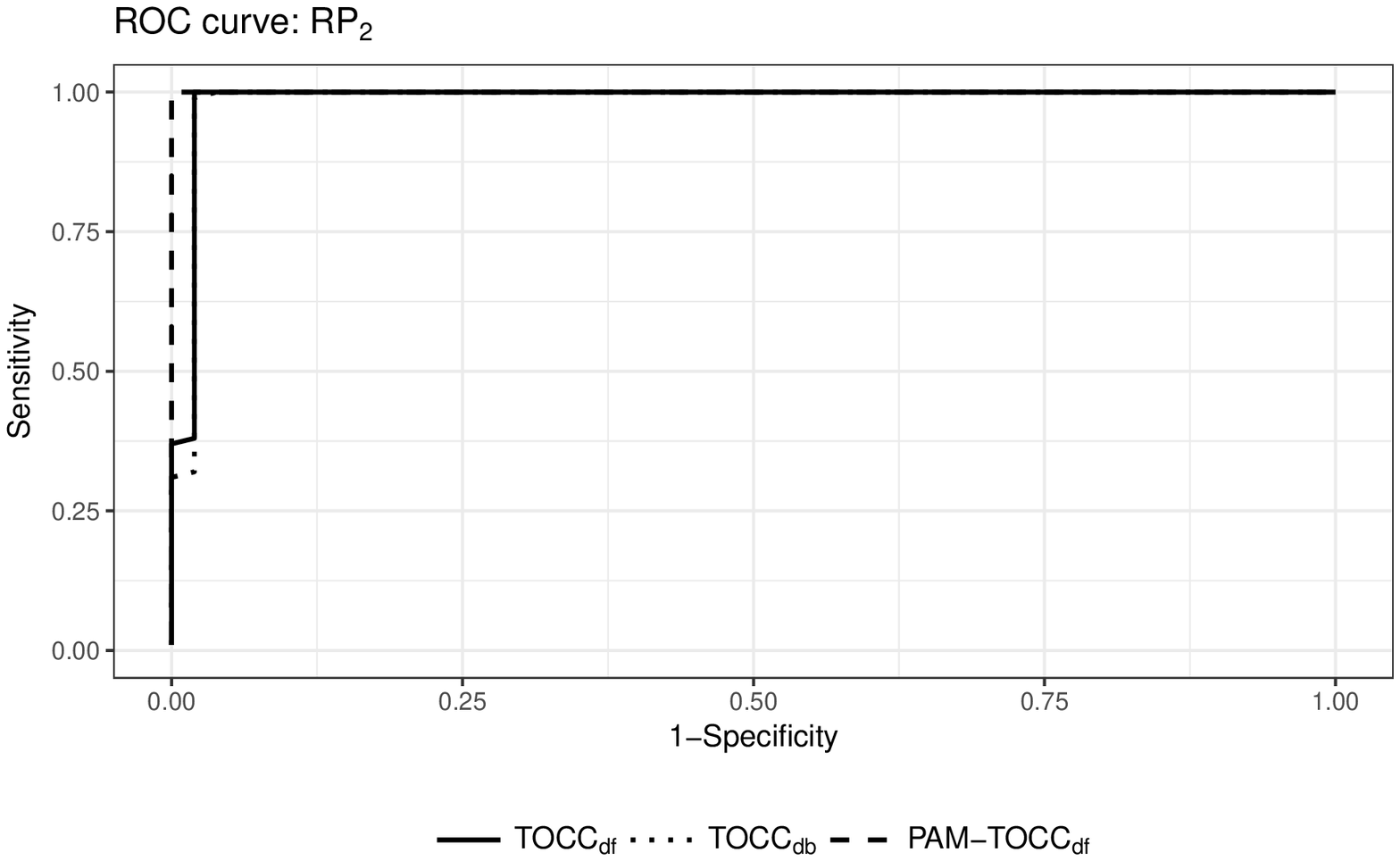}
	\end{subfigure} \\
	\begin{subfigure}[b]{.49\textwidth}     \centering
		\includegraphics[width=\textwidth]{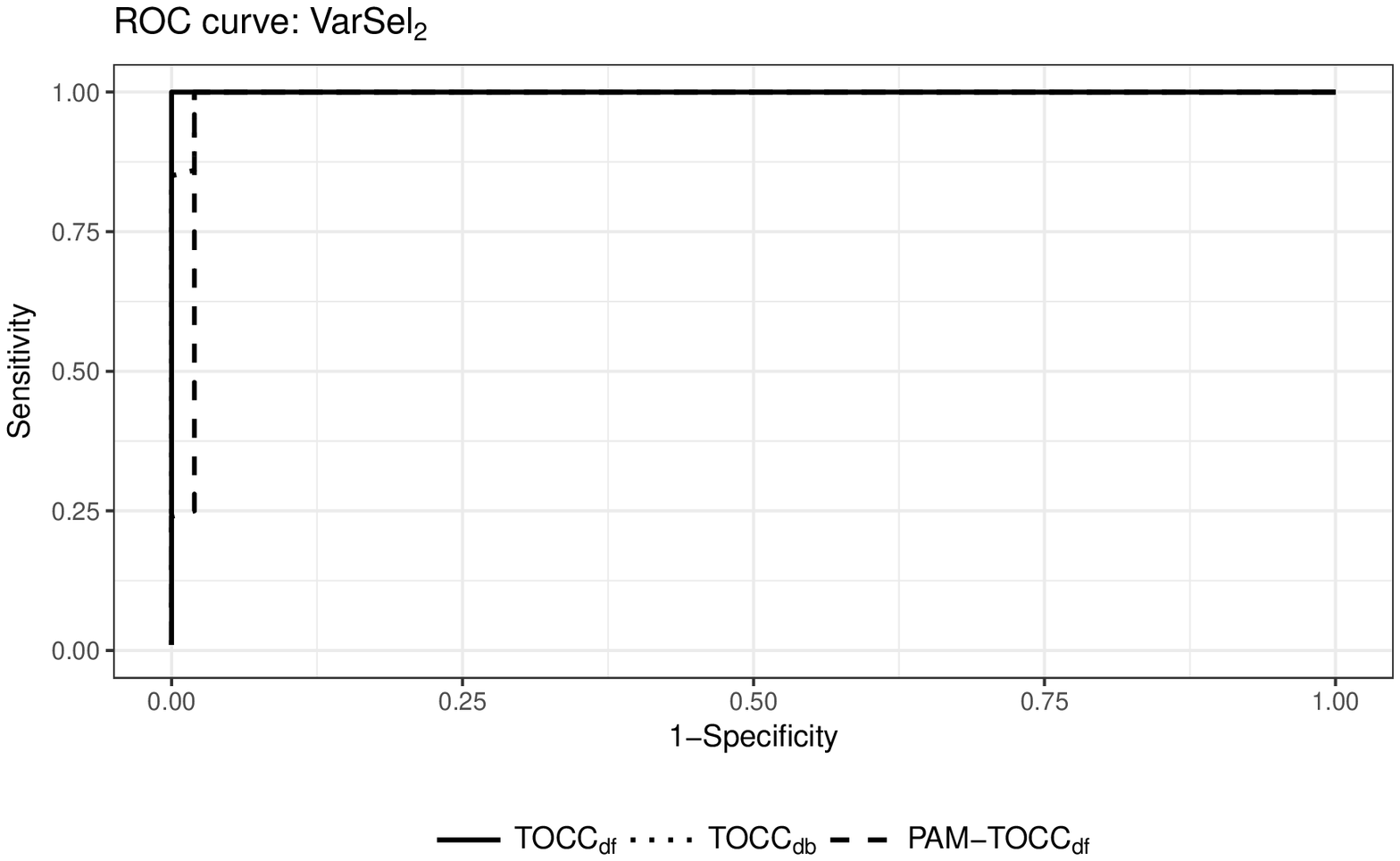}
	\end{subfigure}
	\begin{subfigure}[b]{.49\textwidth} \centering
		\includegraphics[width=\textwidth]{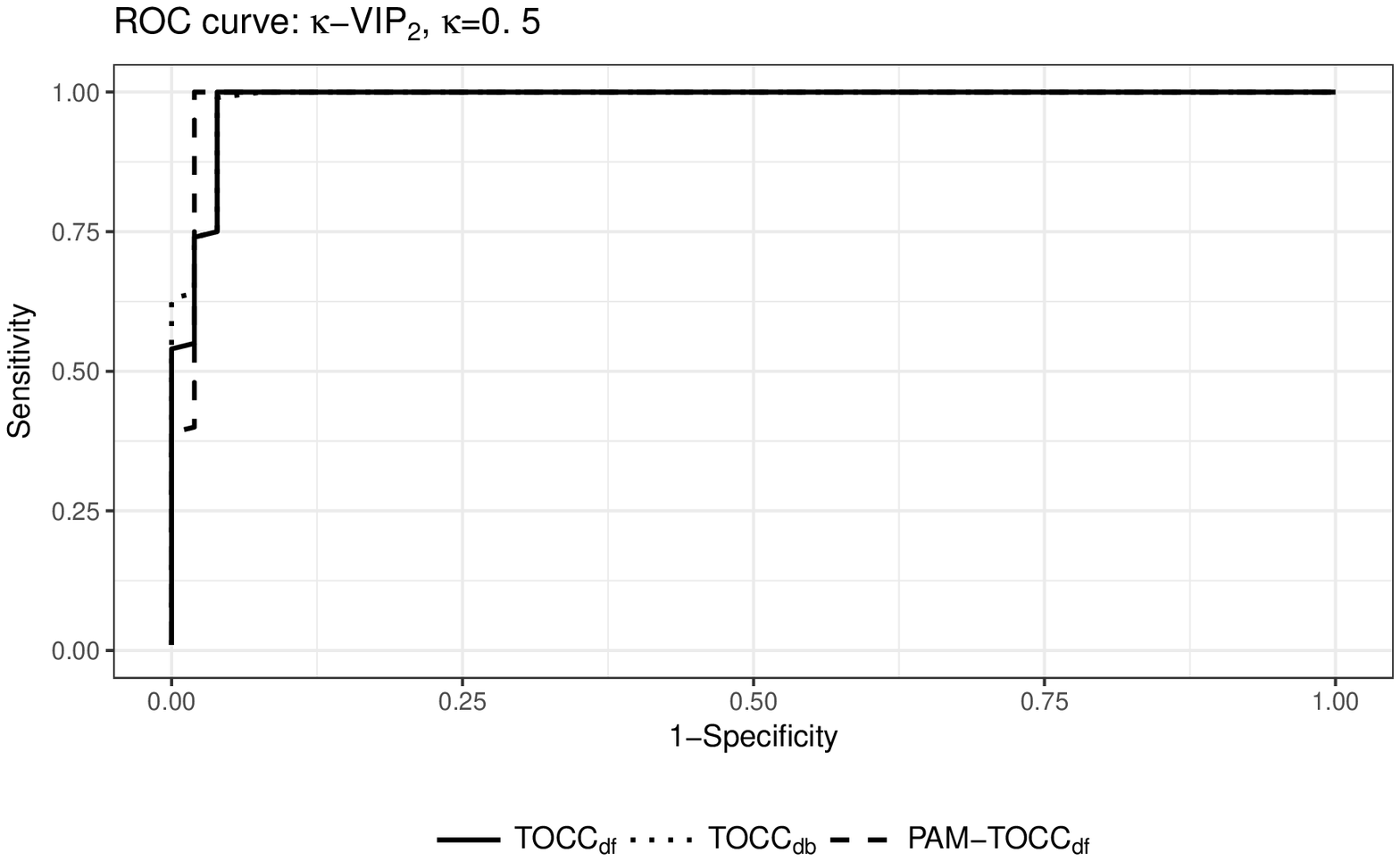}
	\end{subfigure}
	\caption{Glass data: ROC curves of the proposals, distinguished by the different strategies implemented to reduce the data dimensionality.}\label{fig:roc}
\end{figure}

\begin{figure} \centering
	\begin{subfigure}[b]{.49\textwidth}     \centering
		\includegraphics[width=\textwidth]{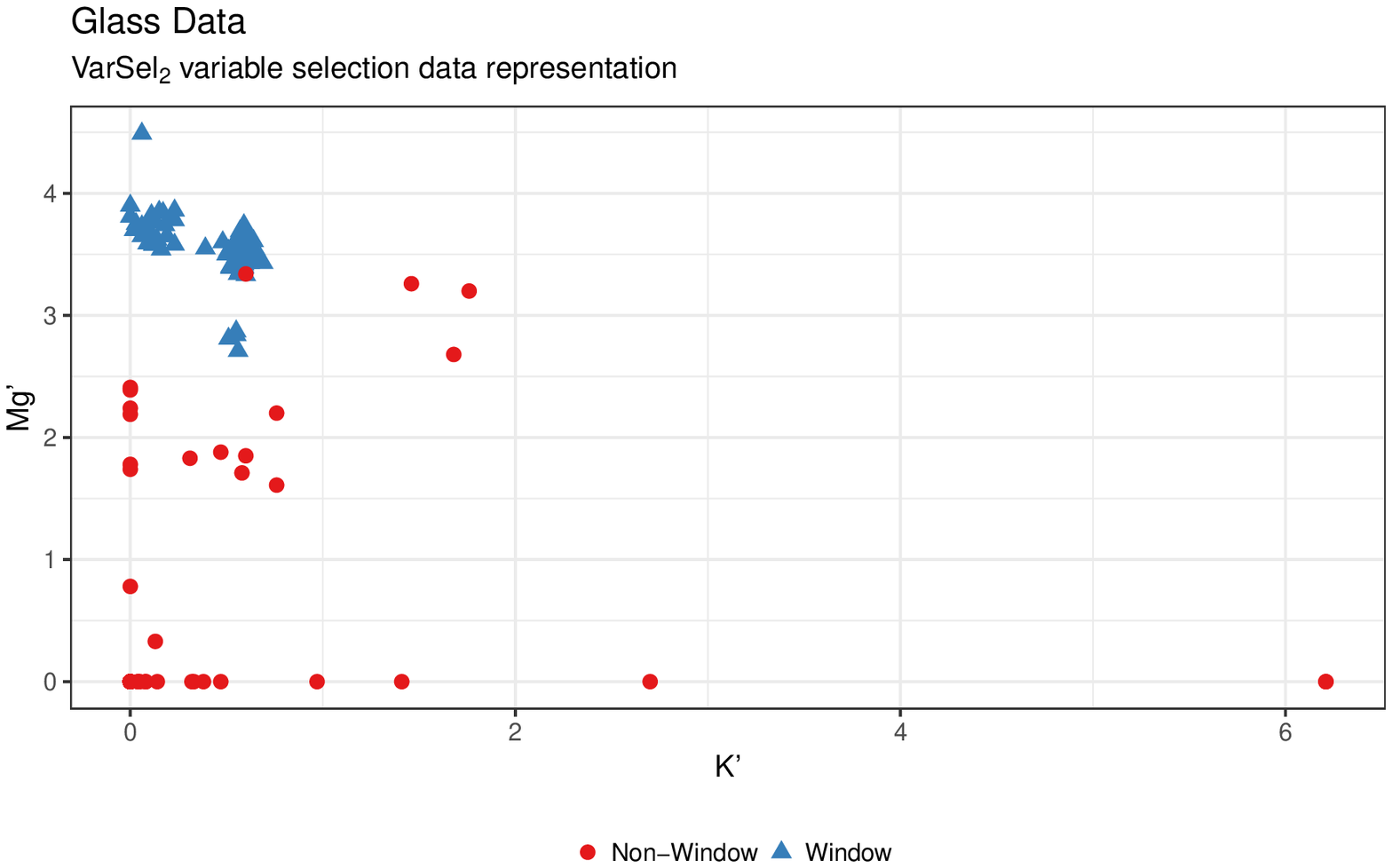}
	\end{subfigure}
	\begin{subfigure}[b]{.49\textwidth}     \centering
		\includegraphics[width=\textwidth]{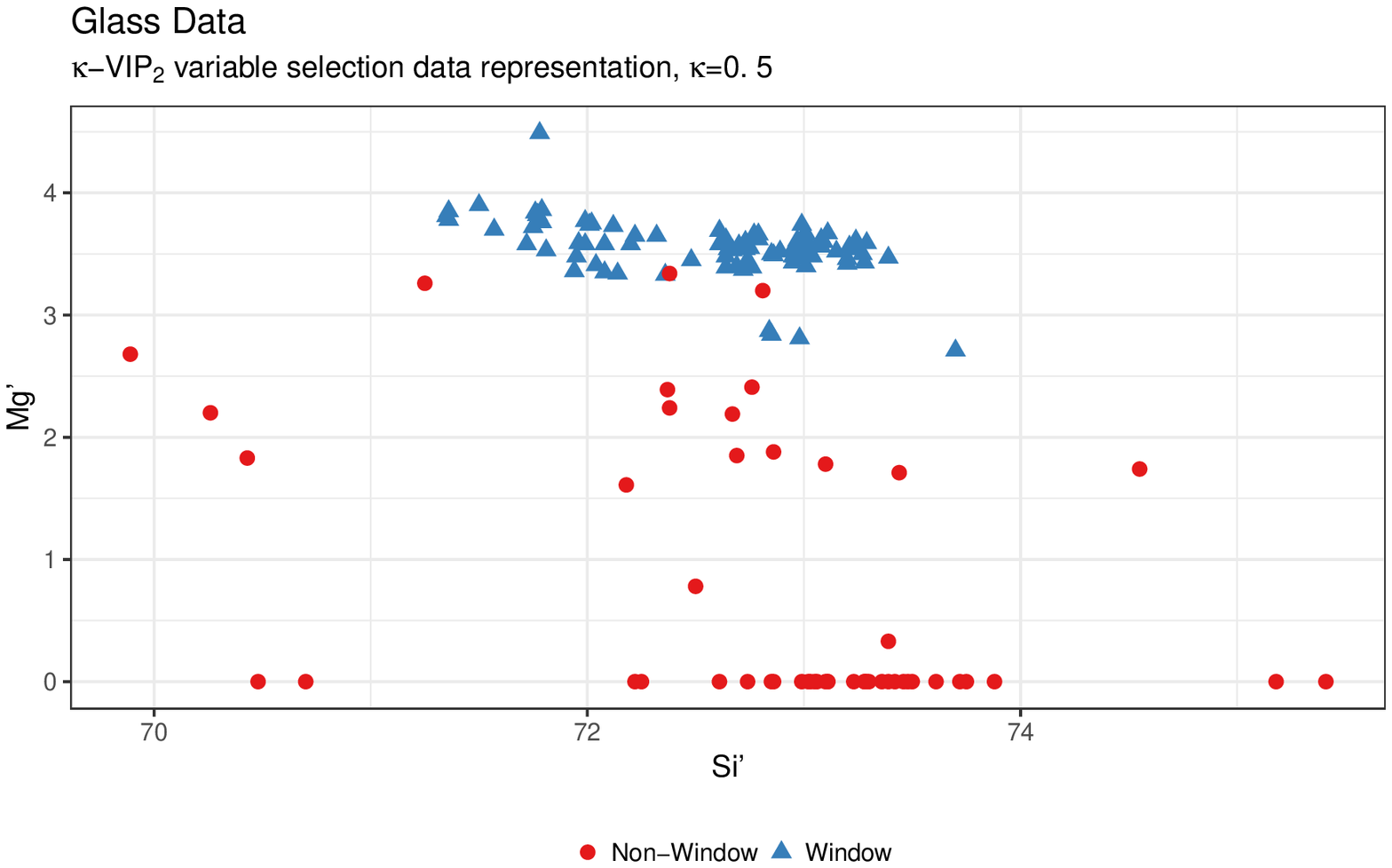}
	\end{subfigure} \\
	\caption{Glass data: bi-dimensional data representation according to the variable selection procedures.}\label{fig:varsel}
\end{figure}

\begin{table}
    \centering
    \caption{Glass data: area under the ROC curve (AUC). The subscript below each dimension reduction or variable selection procedure refers to the dimension of the feature space used. $\kappa=0.5$}\label{tab:AUC}
  \fbox{\begin{tabular}{lcccc}
  &  \multicolumn{4}{c}{AUC}\\
\cline{2-5}
                	& PCA$_2$   & RP$_2$    & \emph{varSel}$_2$ & $\kappa$-VIP$_2$   \\ \hline
  TOCC$_{df}$       		& 0.946     	& 0.988     & 1.000                     & 0.986 \\
  TOCC$_{db}$       		& 0.905     	& 0.987     & 0.997                     & 0.988 \\
  PAM-TOCC$_{df}$   	& 0.963     	& 1.000     & 0.985                     & 0.988 \\
  \end{tabular}}
\end{table}

\medskip

Figure \ref{fig:roc} depicts the ROC curves for the three TOCCs, distinguished by the different strategies implemented to reduce the data dimensionality; Table \ref{tab:AUC} contains the corresponding area under the ROC curve (AUC). Overall results are very good, as almost all the non-window fragments have been recognised. However, a few considerations can still be made. In particular, for this set of data variable selection procedures slightly outperform the dimension reduction ones; plots in the second row exhibit a quasi-perfect performance. As shown in Figure \ref{fig:varsel}, the two sets of fragments look well separated when plotted according to the most relevant features, even if these are different for the two methods (\emph{varSel} chose potassium and magnesium, whilst $\kappa$-VIP selected silicon and magnesium). The goodness of such selections allows all the TOCCs to perform excellently.

When the characteristics of the target and non-target objects are not so easily distinguishable (see, Figure \ref{fig:glassdata}), the PAM-TOCC$_{df}$ should be preferred; this method is, by construction, more capable to identify the non-window glasses scattered within the window samples; in addition, it requires the lowest computational time, as shown in Table \ref{tab:spectime}.


\begin{table}
    \centering
    \caption{Glass data: specificity rates corresponding to a sensitivity level $s \ge 0.9$ and corresponding computational time (in seconds). The subscript below each dimension reduction or variable selection procedure refers to the dimension of the feature space used. $\kappa=0.5$.}\label{tab:spectime}
 \fbox{ \begin{tabular}{lccccccccc}
&  \multicolumn{4}{c}{Specificity} && \multicolumn{4}{c}{Time}\\
\cline{2-5} \cline{7-10}
                    & PCA$_2$   & RP$_2$    & \emph{varSel}$_2$ & $\kappa$-VIP$_2$ & & PCA$_2$   & RP$_2$    & \emph{varSel}$_2$ & $\kappa$-VIP$_2$  \\ \hline
  TOCC$_{df}$       & 0.882     & 0.980     & 1.000             & 0.961         & &  0.23     & 7.19     & 0.09             & 0.08  \\
  TOCC$_{db}$       & 0.804     & 0.980     & 0.980             & 0.961         & & 1.19     & 121.94   & 1.19             & 1.43  \\
  PAM-TOCC$_{df}$   & 0.922     & 1.000     & 0.980             & 0.980       &  & 0.09     & 2.30     & 0.04             & 0.03 \\
  \end{tabular}}
\end{table}
%

\section{Discussion and conclusions}

In this work, new directions for forensic analysis of glass fragments have been considered. In particular, the problem of identifying glass samples that come from different sources in a crime scene has been addressed for the first time (to the best of our knowledge) within a one-class classification framework.

We proposed to consider \emph{transvariation probability} as a measure of resemblance between an observation and a set of well-known objects. Basing on \emph{tp}, three different algorithms have been introduced, according to the available information on the target set. Namely, TOCC$_{df}$ is a distribution-free method that only relies on the computation of the transvariation probability. When information on the distributional shape of the target units is available, a distribution-based TOCC, TOCC$_{db}$, can be successfully implemented. These methods perform very well, especially when non-target objects lie on the external perimeter of the target class.

%

However, information on the deviating samples is, in principle, not available and the situation just described may not be realistic as non-target units can actually pollute the target set intrinsically.
For this reason, a more flexible method that allows to \emph{peel} the target objects within the data cloud has been developed.
The PAM-TOCC$_{df}$ identifies homogeneous groups of target samples and exploits such information to spot the units that deviate from each cluster.

The performances of the proposed method have been evaluated in terms of specificity, i.e. the proportion of actual negatives that are correctly predicted, on multiple synthetic datasets. Simulation results demonstrate that the use of $tp$ as a tool for one-class classification outperforms several state-of-the-art methods.

The chemical composition of the two sets of glass fragments that motivate our work is very similar and the samples cannot be easily distinguished. For this reason, the PAM-TOCC$_{df}$ appears to be the most appropriate transvariation-based one-class classifier, being able to detect all the non-window objects.
The methodology we propose is very flexible and can be employed to solve different one-class classification tasks,
such as food authentication, fraud detection, central statistical monitoring issues, to name a few.
In \citep{amsdottorato8412} excellent performances achieved by the TOCCs on other datasets are shown.
In particular, the proposed classifier has been applied to two sets of near infrared spectroscopic food data,
in order to evaluate food samples' authenticity (namely, one related to honey samples and the other concerning olive oil).
In addition, the Water Treatment Plant dataset from the UCI repository was successfully explored in a fault detection perspective.
This dataset is well-known in the literature as a difficult classification task, since no method
turned out to be able to correctly identify the days in which the plant wrongly operated.

\bibliographystyle{plain}

\bibliography{oneclassbib}
\end{document}